\documentclass[manuscript,showpacs,amsmath,amssymb,nofootinbibtwocolumn]{revtex4-1}
\usepackage{amsmath}
\usepackage{amssymb}
\usepackage{graphicx}
\usepackage{dcolumn}
\usepackage{bm}

\usepackage{xcolor}

\begin{document}
\title{Optimal Noise in a Stochastic Model for Local Search}
\author{J. Noetel$^1$, V. L. S. Freitas$^2$, E. E. N. Macau$^{2,3}$, L. Schimansky-Geier$^{1,4}$}
\affiliation{$^1$Institute of Physics, Humboldt University at Berlin,
  Newtonstr. 15, D-12489 Berlin, Germany\\
  $^2$National Institute for Space Research 12227-010 Sao Jose dos Campos, Brazil\\$^3$Federal University of Sao Paulo
12247-014 Sao Jose dos Campos, Brazil\\$^4$Berlin Bernstein Center for Computational Neuroscience, Humboldt University at Berlin, Unter den Linden 6, D-10099 Berlin, Germany  }

\begin{abstract}
We develop a prototypical stochastic model for local search around a given home. The stochastic dynamic model is motivated by experimental findings of the motion of a fruit fly around a given spot of food but shall generally describe local search behavior. 
The local search consists of a sequence of two epochs. In the first the searcher explores new space around the home  
whereas it returns to the home during the second epoch. In the proposed two dimensional model both tasks are described by the same stochastic dynamics. 
The searcher moves with constant speed and its angular dynamics is driven by a symmetric $\alpha$-stable noise source. 
The latter stands for the uncertainty to decide the new direction of motion.  The main ingredient of the model is the nonlinear interaction dynamics of the searcher with its home. In order to determine the new heading direction,  the searcher shall know the actual angles of its position to the home and of the heading vector. A bound state to the home is realized by a permanent switch of a repulsive and attractive forcing of the heading direction from the position direction corresponding to search and return epochs. Our investigation elucidates the analytic tractability of the deterministic and stochastic dynamics. Noise transforms the  conservative deterministic dynamics into a dissipative one of the moments. 
The noise enables a faster finding of a target distinct from the home  with optimal intensity. This optimal situation is related to the noise dependent relaxation time. It is uniquely defined for all $\alpha$ and distinguishes between the stochastic dynamics before and after its value. For times large compared to this we derive the corresponding Smoluchowski equation and find diffusive spreading of searcher in the space. We report on the  qualitative agreement with the experimentally observed spatial distribution, noisy oscillatory return times, spatial autocorrelation function of the fruit fly. 
But as result of its simplicity the model aims to reproduce the local search behavior of other 
units during their exploration of surrounding space and their quasi-periodic return to a home. 
\end{abstract}

\begin{center}
\end{center}

\maketitle

\section{Introduction}
In the present paper we propose a rather general but simple model for a local search. In contrast to global search, 
local search is known to be concerned with the neighborhood of a given spot \cite{Klages}, called home. 
Such locally searching devices, living objects, agents, etc. return permanently to the home. In fact, 
they form a bound state with the home without real acting attractive forces. The behavior of maintaining a reference location
and returning to it is called ``homing''\cite{Mittelstaedt}. If
the position of the home and the angle towards the home 
are known then the method is called path integration
\cite{Mittelstaedt, Cheng, Wang,Kim_Dickinson_2017}. 

The homing behavior might be based on idiothetic (internal), or allothetic (external) cues \cite{Zeil}. 
An external cue might be, for example, the position of the sun, odors or pheromones, or regional landmarks 
while idiothetic cues are based on some internal storage mechanism \cite{Seelig,Green}.

Such homing behavior is also known from insects such as ants, bees and
flies \cite{Vickerstaff,Wehner_1981,Wehner_et_al_1996,ElJundi_2017,Kim_Dickinson_2017}. 
The home can be a nest or a source of food.  

Various objects perform homing in different ways. The ant \textit{Cataglyphis} builds a tortuous trajectory until a source 
of food is found, and then goes back home in a straight path,
indicating it has a rather accurate sense of the nest position \cite{Vickerstaff, Wehner_1981,Wehner_et_al_1996}. In contrast, the 
fruit fly \textit{Drosophila melanogaster} performs idiothetic homing after a food source is found.  The fly explores 
vicinity of the food, returns and starts exploring again. This oscillating behavior is believed to
fulfill the purpose of foraging, so the flies are then able to search
for more food, while keeping track of already found sources
\cite{ElJundi_2017}. In all the mentioned examples of living objects the trajectories always appear stochastic.

Furthermore, a new age of exploration starts to develop, from the technology point of view \cite{Chien_2017}. Explorer robots are being projected 
for missions in the  ocean \cite{Hook_2013,Girdhar_2011,Leonard_et_al_2007} and space \cite{Chien_2017, Dubowsky_2005}. 
Autonomous vehicles \cite{Leonard_et_al_2007,Duarte_et_al_2016,Nirmal, Moeller} will be used for data collection and local search. 
In the development of an scientific understanding of these technologies and their devices, research in this field relates in 
many cases successfully on data and/or numeric models developed for animal navigation \cite{Chien_2017}. Often such devices are inspired by 
research on biological objects fulfilling different purposes, as for example, local search as a permanent search 
and return process. A better theoretic-mathematical understanding of local search, might also explain the performance of  self-navigating objects.


In this paper we introduce a class of minimal stochastic models for local search that does not distinguish between 
the search and the return. Both epochs shall follow the same law. The particular dynamic model and its 
simplicity shall be justified by the allowance for an analytical tractability and that the model reproduces qualitatively 
several  experimental findings. We use a Langevin equation, that considers an active Brownian particle with constant 
speed whose spatial motion is two dimensional. The constant speed is common in a variety  of models \cite{mikhailov,Romanczuk} 
and also observed in \cite{Kim_Dickinson_2017}. The model aims to mimic the motion of simple organisms, 
in particular, we orientate the discussion of the analytic and numeric results of our model on findings for 
the fruit flies considered in \cite{Kim_Dickinson_2017}. 

We implement the local search around the home via a nonlinear coupling term between the heading angle of the particle 
and the angle formed by its current position and the home. Both angles effectively interact in a sequence of an escape 
and pursuit dynamics. To perform this dynamics, we assume that the agent possesses an internal storage mechanism 
for the current angles of the heading direction and of the position towards the home. 
The resulting spatial motion allows the particle to explore the vicinity of the home in order to find new food sources and 
searchers will consecutively return to the home.

In Section \ref{sec:model}, we briefly introduce the model, then discuss in Section \ref{sec:model_det} 
the deterministic trajectories, that turns out to be a conservative system with an integral of motion. We 
continue in Section \ref{sec:model_noise} the discussion with
noise present. The noise will be added to the angular dynamics. In many stochastic models Gaussian white noise is applied. 
However, we model the noise as symmetric $\alpha$-stable white noise source which also includes the case of a Gaussian noise 
with the particular choice $\alpha=2$. 
Thus, we are able to include different turning statistics of the heading angle including non-Gaussian 
white noise as it was observed in the searching motion of fruit flies and for other insects \cite{Garcia_daphnia}. 
For global search the L\'evy Flight Hypothesis \cite{Klages} is popular. However, for local search, step 
lengths distributed according to a power law seems counter productive.  
To our best knowledge we do not know about models for local searchers  with corresponding noise sources.

The noise serves as an uncertainty in the heading direction for the active particle. This
uncertainty might have its origin in a limited capability to choose an exact direction of motion, due to external influences or 
the mechanics of the brain. Another source of randomness is the lack of information. Under the circumstance where it is unclear 
what the optimal heading direction 
is the actual choice can become random. Those uncertainties cause in our model a
steady state distribution (Section \ref{sec:spat_dist}) of the particles centered at the home with 
exponentially decaying probability density at large distances.
In Section \ref{sec:disc} we discuss the local search characteristics and 
especially we find an optimal noise strength for discovering food.
In Section \ref{sec:extension}, we generalize the model through allowing the coupling to the home to be dependent 
on the distance to the home. Finally, we summarize our findings in Section
  \ref{sec:concl}. 
  
Several technical aspects concerning the $\alpha$-stable noise have been included in the Appendices \ref{app:X-dyn} and \ref{app:r}. 
They have been of importance for the investigations on stochastic dynamics driven by $\alpha$-stable noise in 
the angular dynamics. In \ref{app:X-dyn} we give the foundation for the noise dependent relaxation time of the system. 
It serves as the characteristic time for relaxation of angular asymmetry. For larger times, the system is approximated by 
an overdamped description as elucidated in Appendix \ref{app:r}  for various sources of $\alpha$-noise in the angular dynamics.  
Appendix \ref{app:celes} describes the deterministic dynamics as a celestial mechanics with constant speed.

\section{The Model}
\label{sec:model}

\begin{figure}[h]
  \includegraphics[width=0.42\linewidth]{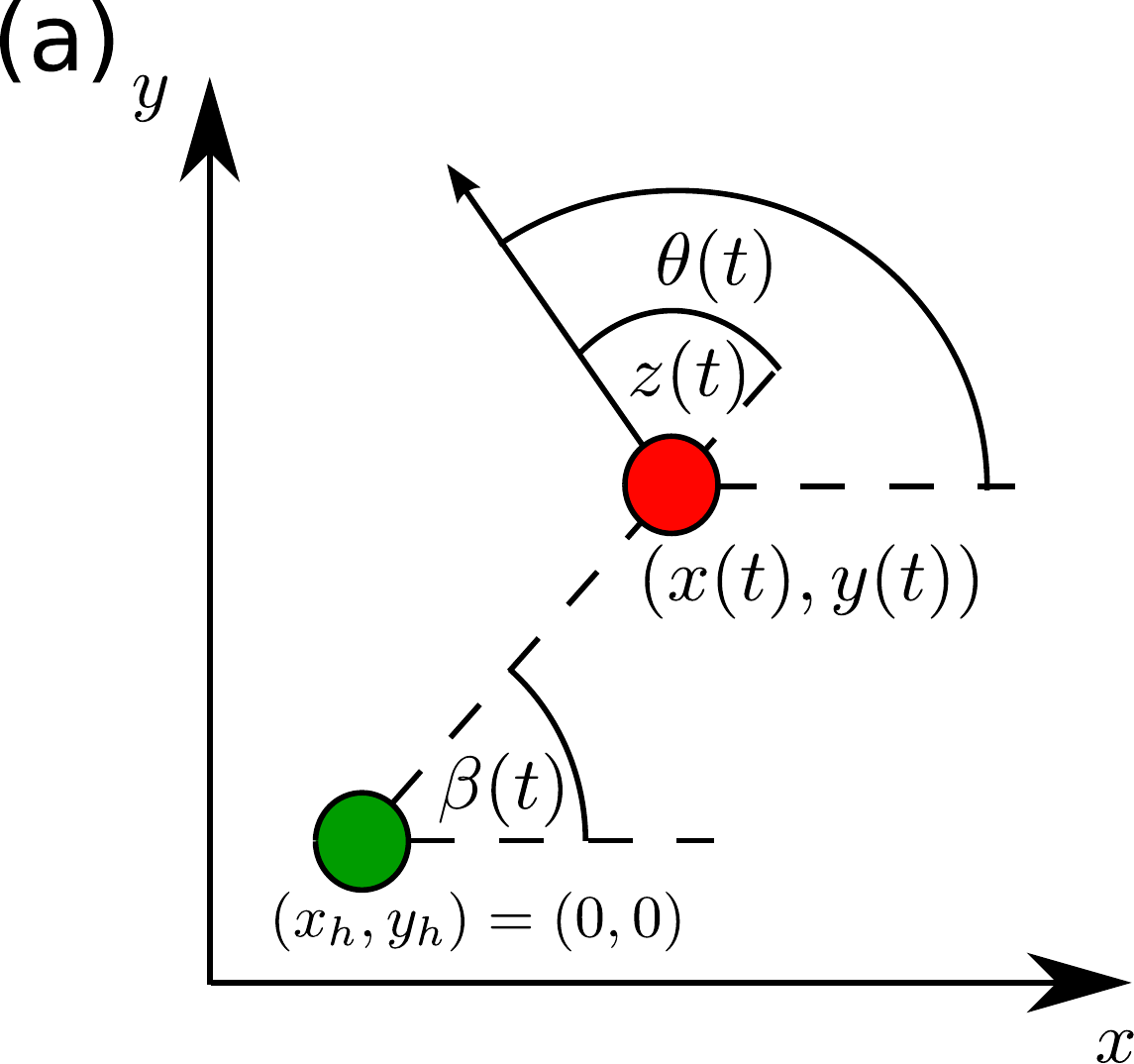}~~~~~~~~~~~~~\includegraphics[width=0.33\linewidth]{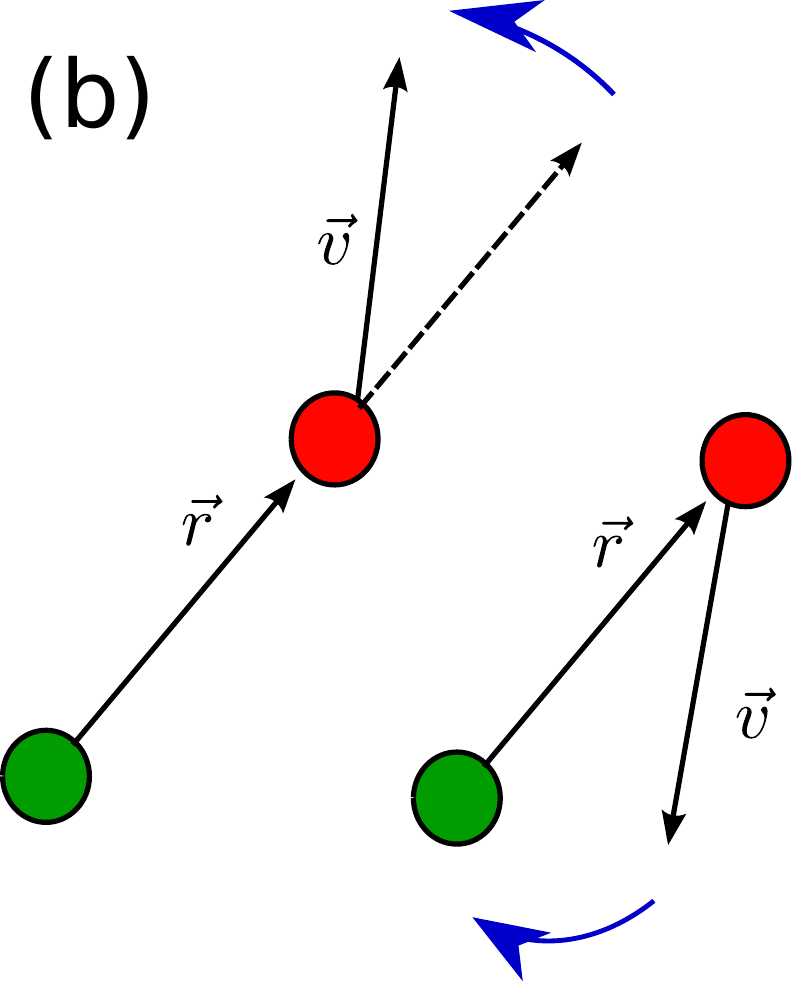}   
  \caption{(a) Schematic representation of coordinates. The angle $\theta(t)$ defines the current heading direction pointing 
  along the actual velocity. 
    The angle $ \beta(t)$ is the direction of the vector from the home to the agent positioned at $\vec{r}(t)$. 
    For convenience the home is situated at the origin $\vec{r}_h=(0,0)$. (b) Sketch of the interaction between position 
    and heading vectors. If the heading vector points outwardly from the home as the position vector always does, 
    the heading vector becomes repelled from the position vector. In contrast, pointing homewardly the heading vector is 
    attracted by the position vector.}
    \label{fig:beta}
\end{figure}

We consider an active particle whose position vector is given by $\vec{r}(t)=\{x(t),y(t)\}$. 
We model a particle with constant speed $v_0$      
\begin{eqnarray} 
\dot{\vec{r}}=\vec{v}(t)=v_0
\begin{pmatrix}
         \cos\theta(t) \\ \sin\theta(t)
        \end{pmatrix}\,,
\label{eq:r_dot}
\end{eqnarray}
$\dot{\vec{r}}$ denotes the temporal derivative of the position vector $\vec{r}(t)$ and $\theta(t)$ is the heading angle of 
the particle, as it is depicted in Figure \ref{fig:beta}. 

As we chose the home to be situated at the origin of the Cartesian reference frame, the position vector $\vec{r}(t)$ points always 
out of the home in direction of the particles current position. 
The corresponding angle $\beta(t) \in[0,2\pi)$ is given by:
\begin{equation}
\beta(t) = \arctan{\frac{y(t)}{x(t)}}, 
\label{eq:beta}
\end{equation}
as also sketched in Figure \ref{fig:beta}. The time evolution of $\beta$ is determined by Equation \eqref{eq:r_dot}.
For particles which arrive at times $t_h$ at the home $\vec{r}(t_h)=\vec{r}_h$,  
the angle $\beta(t_h)$ remains undefined. In these rare cases we will agree  that the angle $\beta$ converges with the heading direction, 
i.e. it holds $\beta(t_h)=\theta(t_h)$. This choice corresponds to a particle that leaves the home in radial direction. 
Notice, that during the passage of the home, the angle $\beta$ jumps by $\pi$.

The search and return dynamics of the active particle is encoded in the evolution of the heading angle $\theta \in[0,2\pi)$. 
The evolution of the heading direction contains the actual decision
process of the searcher by selecting the future direction of its velocity. We assume that the heading evolves in time according to:
\begin{equation}
\dot{\theta} =  \kappa \sin(\theta-\beta) + \frac{\sigma}{v_0} \xi(t).
\label{eq:dottheta}
\end{equation}
While Equation \eqref{eq:r_dot} is simply the mechanics of the motion with constant speed, Equation (\ref{eq:dottheta}) 
expresses the searcher's  wish,  (i) to explore new space around the home,   and (ii) to return sequentially towards the home. 
It is the process for which the searcher needs the knowledge of the two angles  $\beta(t)$ and $\theta(t)$. 
With the help of both the searcher performs 
path integration.

The first term on the right hand side (r.h.s.), the deterministic term, can be motivated by an effective escape and pursuit 
dynamics as it was also discussed in \cite{romanczulsgcouszin2009}. For positive value of $\kappa$ the heading angle escapes 
the unstable outward direction $\theta(t)=\beta(t)$ and pursues the homeward direction $\theta(t)=\beta(t)+\pi$ 
which is to be stable. Such behavior is schematically indicated in Figure \ref{fig:beta} (b). 
If the projection of the heading onto the position vector is positive the heading becomes repelled from the position vector. 
In contrast, in case that the projection of the heading vector is negative or it is anti-parallel to the position vector, 
the heading is attracted by the homeward direction.

The second term on the r.h.s. of Equation \eqref{eq:dottheta}  $\xi(t)$ shall be a symmetric $\alpha$-stable white noise source.  
It serves as an uncertainty in the heading direction, caused by a decision process, by a limited knowledge of the heading direction 
or by external influences. We also point out that lack of information is a noise source. Lack of information can imply that 
a decision whether to turn left or right or move straight forward is a random choice, as it is unclear what would be the ``right'' choice. 
The noise strength is $\sigma$.
The $\alpha$-stable white noise with $\alpha < 2$ in the angular dynamics yields a continuous description for a run and tumble 
like motion with fast tumbling epoch \cite{noetel:2017} as it was also found and  reported in the experimental study \cite{Kim_Dickinson_2017}. 
In the case of $\alpha=2$, increments of the angle are uncorrelated  in time and with Gaussian support.

The equations of motion \eqref{eq:r_dot} and \eqref{eq:dottheta} take in polar coordinates $r(t)=\sqrt{x^2(t)+y^2(t)}$ 
and $\beta(t)$ given by \eqref{eq:beta} 
 an especially simple form:
\begin{equation}
  \dot{r} = v_0 \cos(\theta-\beta),
  \label{eq:dotdist}
\end{equation}
for the radial velocity and
\begin{equation}
  \dot{\beta} =  \frac{v_0}{r} \sin(\theta-\beta), 
\label{eq:dotbeta}
\end{equation}
for the tangential velocity. Although the distance $r(t)$ and the position angle $\beta(t)$ are stochastic values 
their respective equations of motion do not contain noise sources, as the Cartesian velocities also do not contain noise sources. 
Despite the stochastic character the speed of 
the particle is always constant 
i.e. $\dot{r}^2+(r\dot{\beta})^2=v_0^2$.

Defining now the angle  $z(t)\in{(-\pi,\pi]}$ as the difference of the heading and the
position angle, i.e. $z(t)=\theta(t)-\beta(t)$, we derive
\begin{equation}
  \dot{r} = v_0 \cos(z),
\label{eq:dotr}
\end{equation}
\begin{equation}
  \dot{\beta} = \frac{v_0}{r} \sin(z)
\label{eq:dotbeta1}
\end{equation}
and consequently
\begin{equation}
  \dot{z} = \dot{\theta}-\dot{\beta}=\left(\kappa -\frac{v_0}{r}\right) \sin(z)+\frac{\sigma}{v_0}\xi(t). 
\label{eq:z_stoch}
\end{equation}
These three equations are the stochastic nonlinear dynamics of the local searcher in polar coordinates. One immediately notices 
that the $\beta(t)$-dynamics \eqref{eq:dotbeta1} separates from the stochastic motion on the $(r,z)$ plane. 
The solutions of \eqref{eq:dotr} and \eqref{eq:z_stoch} determine after insertion in \eqref{eq:dotbeta1} and its integration the 
further behavior of the angle $\beta(t)$.

We also point out that in Equation \eqref{eq:z_stoch} a length scale $r_c$ has emerged for the first time in the description. It reads
\begin{equation}
  r_c=v_0/\kappa\,, 
\label{eq:r_c}
\end{equation} 
and defines the relative angular speed between the position vector and the heading vector. In the noise free setting $(\sigma=0)$ 
both vectors rotate always either clockwise or anti-clockwise. But for distances smaller $r_c$ the heading rotates slower then 
the position vector. Otherwise the heading is faster. The search and return motion will be an oscillatory sequence 
around $r_c$ defined by the interplay of the two vectors.

\section{The deterministic model}
\label{sec:model_det}
We will now discuss the deterministic motion of our searcher. Without noise the time evolution \eqref{eq:z_stoch} of the angle $z$ becomes: 
\begin{equation}
  \dot{z} = \left(\kappa -\frac{v_0}{r}\right) \sin(z)\,. 
\label{eq:dotz}
\end{equation}
Equations \eqref{eq:dotr}, \eqref{eq:dotbeta1} and \eqref{eq:dotz}
define the dynamics of our deterministic searcher and have to be supplemented by
initial conditions $r_0=r(t=0),z_0=z(t=0), \beta_0=\beta(t=0)$.  


\subsection{Fixpoints and separatrices}
The simplest solution in the $(r,z)$ plane are stationary fixed points at $z_0=\pm\pi/2$ and $r_0=r_c$, 
with $\dot{r}=\dot{z}=0$. These fixed points are of the center type and have purely imaginary eigenvalues $\lambda$.  
In detail, these eigenvalues are equal for both centers
$\lambda=\pm {\rm i}\kappa$, with $\kappa$ being the coupling strength
from Equation \eqref{eq:dottheta}. The value $\kappa$ yields also the
circulating frequency of the $\beta(t)$ dynamics in the $(x,y)$ plane.
It is in dependence on the initial sign of $z_0$ either
clockwise or anticlockwise. This solution corresponds to a circular trajectory
in the $(x,y)$ plane with radius $r_0$. The initial heading is perpendicular to the
position vector and it remains so, further on.

The second class of trivial solutions belong to states $z=0,\pi$. Since $z(t)$ is periodic in $2\pi$, 
the latter state is also equivalent to $z=-\pi$. States with $z(t)=0$ define the radially unbounded case. 
At arbitrary distances $r$ the particle moves radially away since $\theta(t)=\beta(t)$. In the noise free 
situation solutions with $z=0$ do never change since $\dot{z}=0$. So the trajectories become straight lines and 
the distance to the home grows unbounded. The particles escape with velocity $v_0$ to infinity 
\begin{eqnarray} 
&&r(t)=v_0 t+r_0\nonumber\\
&&\beta(t)=\beta_0\nonumber\\
&&z(t)=0\,.
\label{eq:r_unb}
\end{eqnarray}

Deterministically, the state $z=0$ can be approached only if either $z_0=0$ or $z_0= \pi$. In the latter case the particle 
started with $z=\pi$ at $r_0>0$ will radially approach the home $z=\pi$, 
pass it while the angle $z$ jumps to $z=0$, as the position angle jumps $\beta=\beta_0+\pi$ and then moves away along with $z=0$.

The distance $r=0$ possesses the meaning of a repelling boundary. It can be approached by $z=\pi$. The state at $r=0,z=\pi$ is 
immediately left and the angle $z$ flips to $z=0$. This dynamic behavior is in agreement with the trajectories running nearby this 
boundary which are unable to attach to the boundary. Oppositely, the $z(t)$ dynamics becomes unlimited fast near $r=0$ as it follows 
from \eqref{eq:dotz} due to  the item $\propto 1/r$ on the r.h.s. In consequence the state $r=0$ plays the role of an extended 
saddle point with an incoming separatrix at $z=\pi$ and an outgoing one along $z=0$. This is also confirmed by the fact that the 
two lines play the role of a separatrix. Thus, both lines divide the $(r,z)$ space into two half planes on which either a 
clockwise or and anticlockwise periodic motion takes place. Both sets of solutions do never merge.
  
  \subsection{The oscillatory $(r,z)$ dynamics}
\label{ssec:model_det_rz}
Other solutions correspond to bounded trajectories in the $(r,z)$ space and possess a maximal and a minimal 
distance from the home. The motion becomes a repetitive change of attraction and 
repulsion towards the home pointing direction that defines an oscillatory motion in the $(r,z)$ plane.

\begin{figure}[h]
    \includegraphics[width=0.55\linewidth]{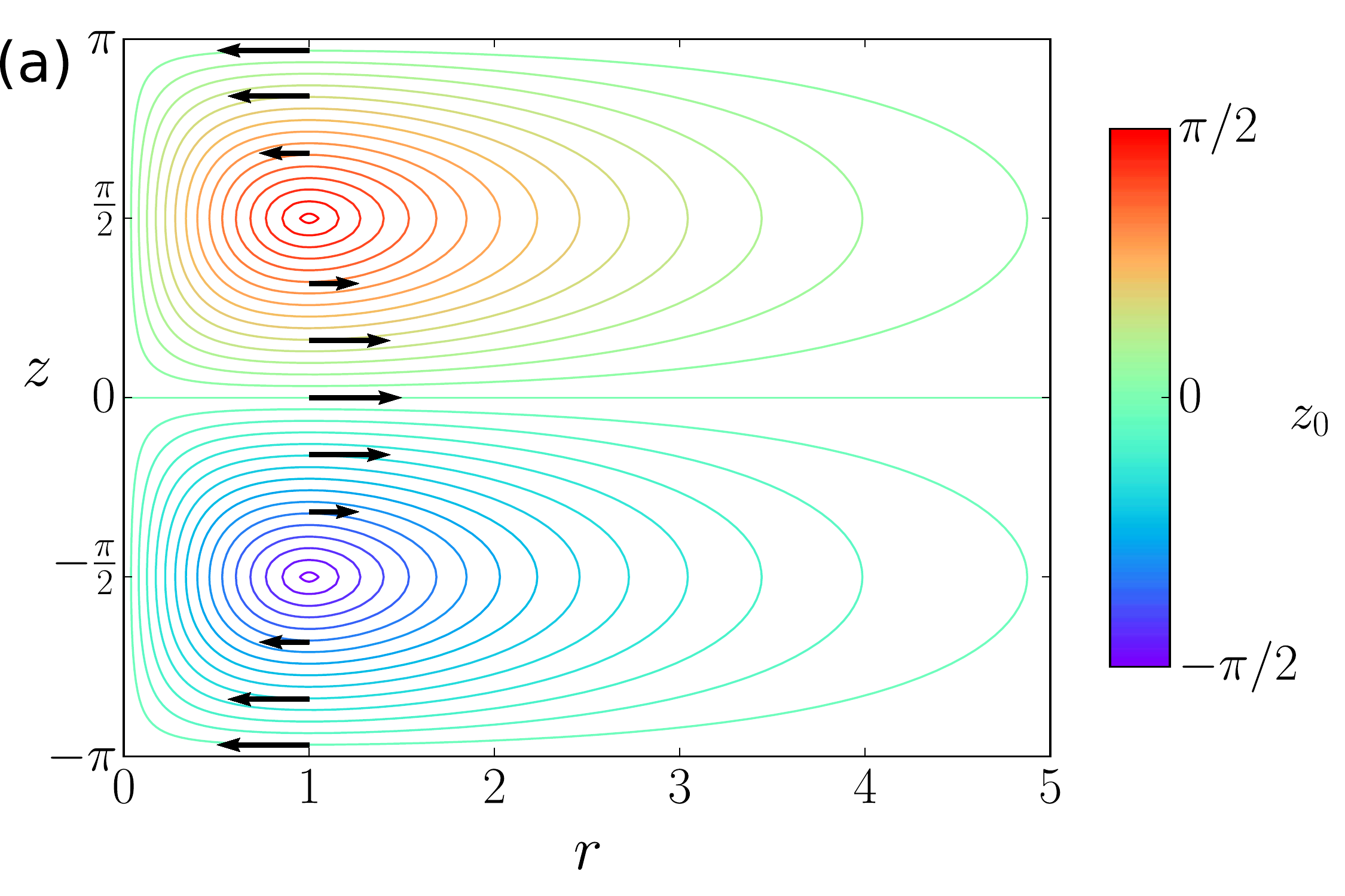}
    \includegraphics[width=0.43\linewidth]{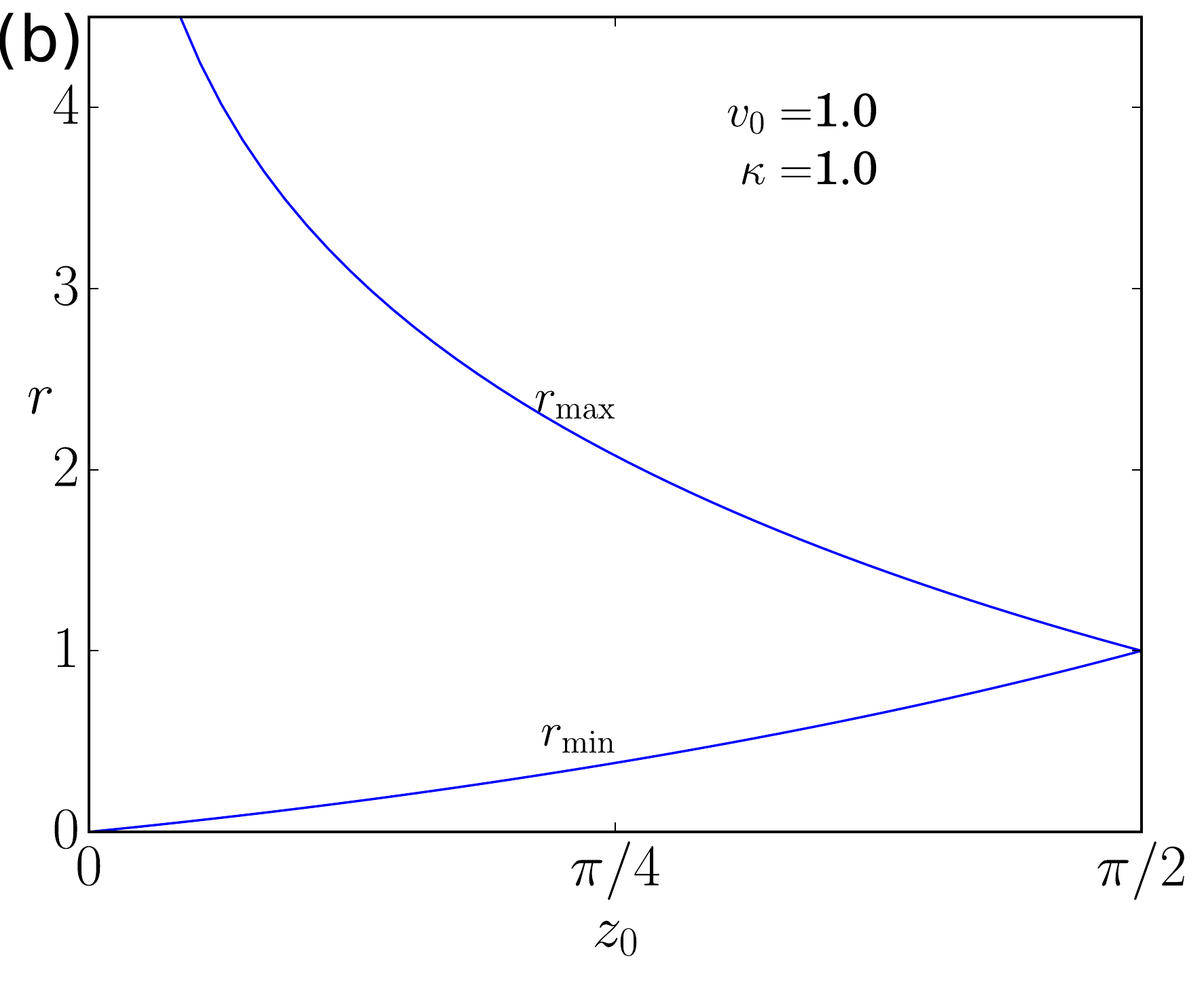}
    \caption{(a) Sample trajectories in $(r,z)$ plane according to Equation \eqref{eq:sinzvr}. 
    Color-bar corresponds to initial starting angle $z_0$ 
    and thus colored trajectories belong also to different values of $X(z_0)$.    
    (b) minimal and maximal distance in dependence of the initial angle $z_0$ for $r_0=r_c$. 
    Parameters: $v_0=1$, $\kappa=1$. }
    \label{fig:r_z_space}
\end{figure}

Figure \ref{fig:r_z_space} presents the flow diagram of the $(r,z)$ dynamics.  
It is obtained when eliminating the time dependence by differentiating the angle $z$ with respect to the position $r$ as
\begin{equation}
\frac{dz}{dr}=\frac{\dot{z}}{\dot{r}} =-\left(\frac{1}{r}-\frac{1}{r_c}\right)\tan(z). 
\label{eq:rofz}
\end{equation}

Since distance $r$ and angle $z$ can be factorized in the differential
Equation \eqref{eq:rofz} we can integrate and find parametrically
$z(r)$. With the initial conditions $r_0$ and $z_0$, we can formally
write the angle in dependence of the position as
\begin{equation}
\sin(z(r)) \exp\left(-\frac{r}{r_c}\right) r= X(z_0,r_0)= \sin(z_0) \exp\left(-\frac{r_0}{r_c}\right) r_0 ={\rm const}.                 
\label{eq:sinzvr}
\end{equation}
This expression defines the formal solution $z(r)$. The parameter $X$
is an integral of motion for the $(r,z)$ dynamics. Therefore, the $(r,z)$ dynamics is conservative. The value of $X$
depends on the initial coordinates, only.  In particular, the sign of
the initial angle also fixes the sign of $z(t)$ during the motion.

In the $(r,z)$ plane, the solutions are periodic, so we can choose the initial condition $r_0=v_0/\kappa=r_c$, as every trajectory 
at least crosses $r=r_c$. Hence, the integral of motion becomes a function of $z_0$, only, i.e. $X=X(z_0)$. This angle $z_0$ can 
be restricted to $z_0 \in [-\pi/2, \pi/2]$.
The motion along $z_0=0$ corresponds to $X=0$. The fixed points have $z_0=\pm\pi/2$ with the $X$-value according to \eqref{eq:sinzvr}.

Figure \ref{fig:r_z_space} (a) shows solutions of
\eqref{eq:sinzvr} in the $(r,z)$ plane, with parameters $v_0=1$,
$\kappa=1$ and $r_0=r_c$. The initial angle $z_0$ is varied corresponding to
the color-bar. As already outlined, there is no trajectory which
crosses the separatrices $z=0$ or $z=\pm\pi$, in fact the angle $z(t)$ is bounded when
considering a specific trajectory. This means that the angular
momentum 
\begin{equation}
L=r^2\dot{\beta}=v_0r\sin(z)=v_0\exp\left(\frac{r}{r_c} \right) X(z_0),
\label{eq:L}
\end{equation}
along a trajectory never changes its sign. It follows that in the
deterministic model the particles either move  with $z>0$ in a clockwise,
or with $z<0$ in a counterclockwise fashion around the home in the Cartesian
coordinate system. The upper and lower half plane in
Figure \ref{fig:r_z_space} (a) correspond to particles starting with either $z_0> 0$ or $z_0<0$, respectively.

Trajectories in the $(r,z)$ plane approach the maximal $r_{\rm max}$ and minimal $r_{\rm min}$ distances (perihelion and aphelion) 
if ${\rm d}r/{\rm d}z$ vanishes. 
To formulate a criterion for these positions and following Equation \eqref{eq:rofz} we set $z=\pm \pi/2$ in 
Equation \eqref{eq:sinzvr} and obtain for both distances 

\begin{equation}
{r_{\rm max/min}}\exp\left(-\frac{\kappa}{v_0}r_{\rm max/min}\right)= |X(z_0)|.
\label{eq:extreme}
\end{equation}
The extremal distances in dependence of the initial angle $z_0$ for $r_0=r_c$ are plotted in Figure \ref{fig:r_z_space} (b) for the
parameters $v_0=1$, $\kappa=1$.
At the extremal distances the radial velocity $\dot{r}$ vanishes. This can be seen when considering the kinetic energy of the 
constant speed particle,
given by $E_{\rm kin}=(\dot{x}^2+\dot{y}^2)/2=v_0^2/2$. In polar coordinates the energy reads 
$E_{\rm kin}=(\dot{r}^2+r^2\dot{\beta}^2)/2$. We express the energy as
\begin{equation}
E_{\rm kin}=\frac{1}{2}\dot{r}^2+\frac{1}{2}v_0^2\sin^2(z(r))=\frac{1}{2}v_0^2
\label{eq:ekin}
\end{equation}
through \eqref{eq:dotbeta1}. For the radial velocity follows:
\begin{equation}
\dot{r}^2=v_0^2\left(1-\sin^2(z(r))\right).
\label{eq:radvel}
\end{equation}

At $r_{\rm max/min}$ the radial velocity vanishes $\dot{r}(r=r_{\rm max/min})=0$. Those are turning points, the motion away from the home turns to motion towards the home and vice versa. 
The corresponding tangential velocity is maximal $\dot{\beta}(r=r_{\rm max/min})=\pm v_0$. 
At $r=r_c$ the radial velocity is maximal, as $0=d\dot{r}/dr=v_0\sin(z)\tan(z)(1/r-1/r_c)$, is solved by $r_c$.
The maximal radial velocity is given by $\dot{r}(r=r_c)=\pm v_0\sqrt{1-\sin^2(z_0)}$, under the condition that 
we set $r_0=v_0/\kappa$. The tangential velocity is therefore given by $\dot{\beta}(r=r_c)=v_0\sin(z_0)$.  
\subsection{Period length and shift of the position angle}
\label{ssec:period_shift}
As the $(r,z)$ dynamics is periodic, we can determine a period length.
\begin{figure}[h]
  \includegraphics[width=0.45\linewidth]{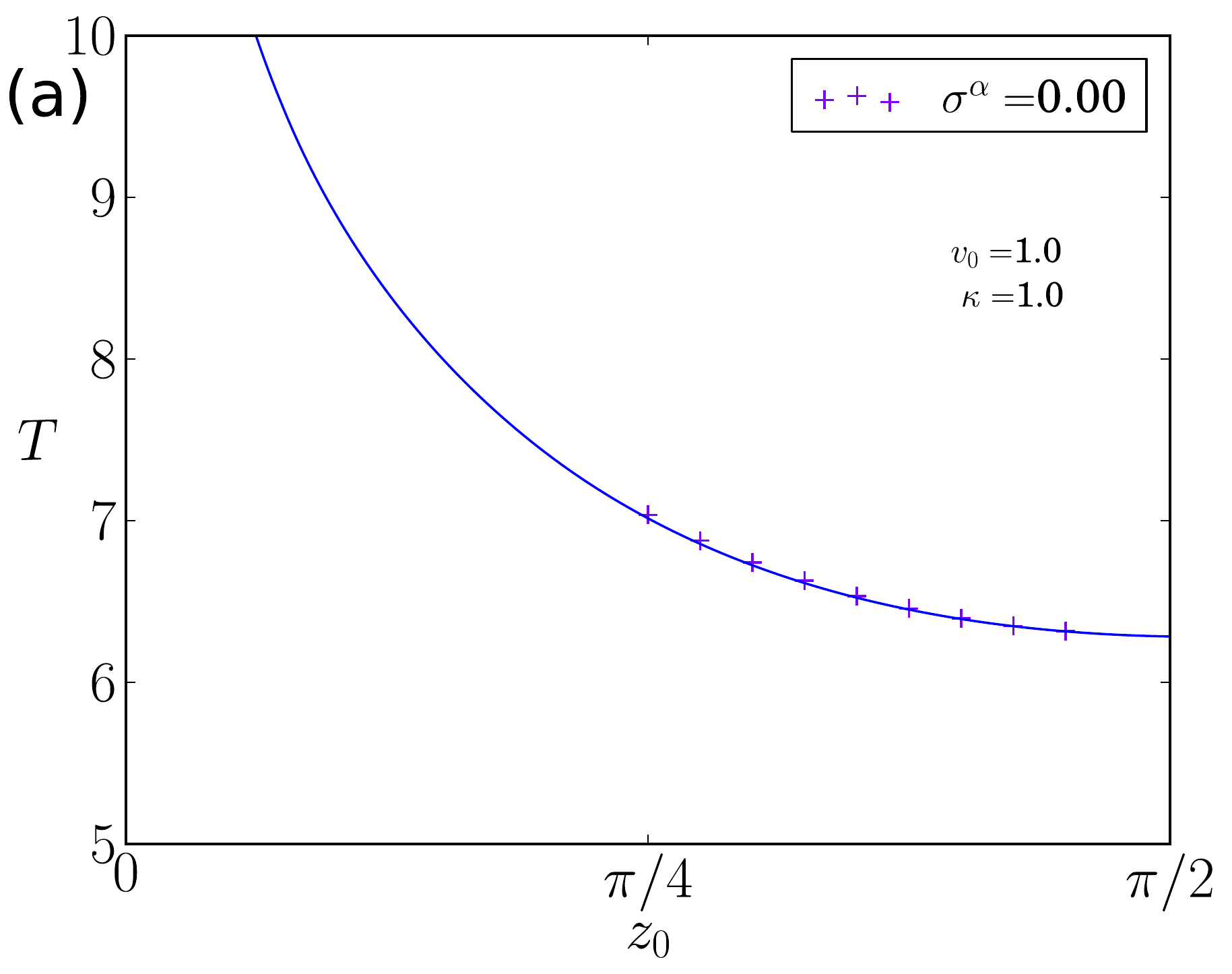}
  \includegraphics[width=0.49\linewidth]{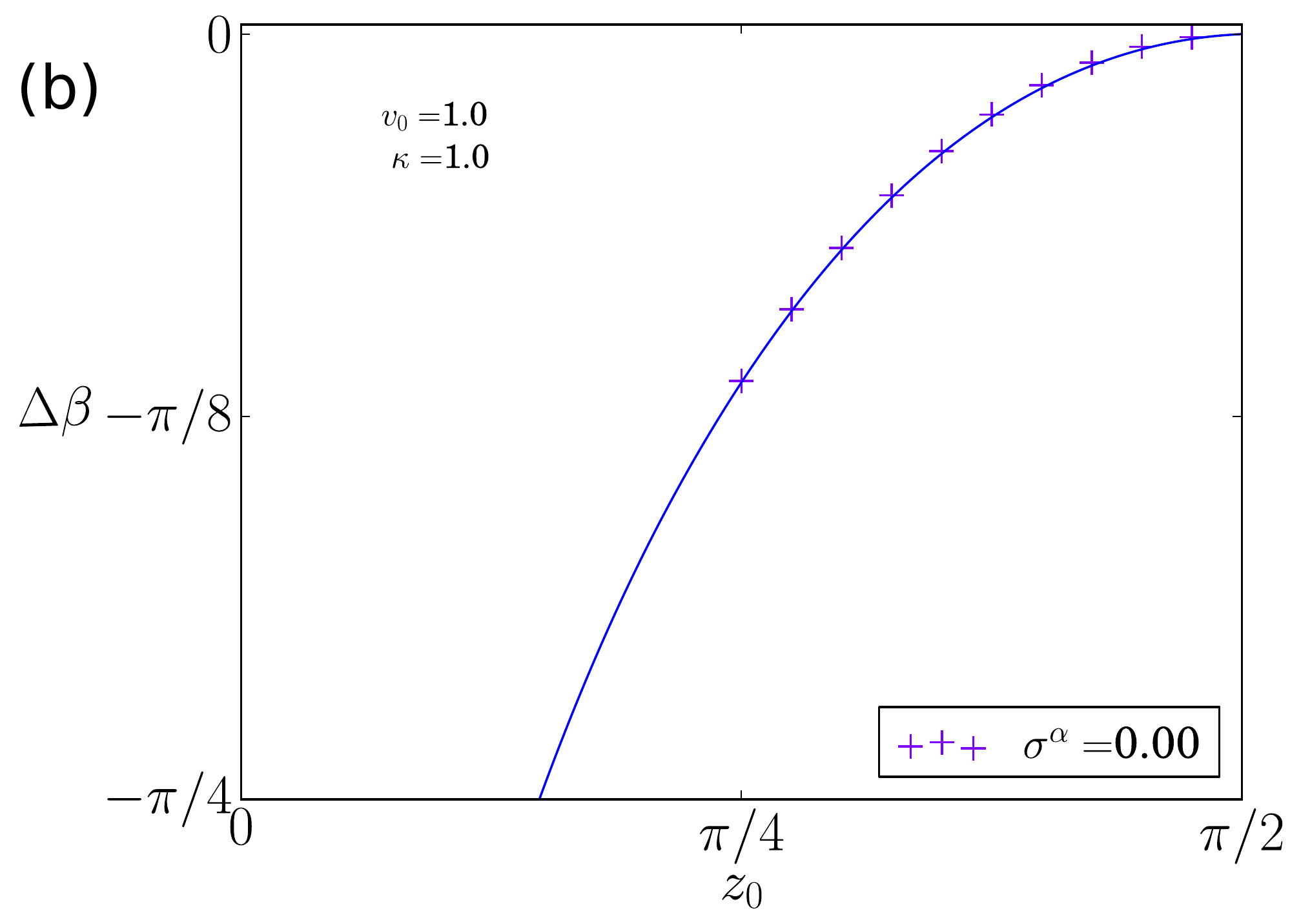}
  \caption{(a) Period of one cycle in dependence of the initial
    angle $z_0$. The solid line corresponds to Equation
    \eqref{eq:period}.  Symbols from simulations of Equations
    \eqref{eq:r_dot} and \eqref{eq:dottheta} with vanishing noise
    ($\sigma=0$).  (b) Value of the apsidal precession of the orbit
    during one revolution as line corresponding to Equation
    \eqref{eq:deltabeta}. Symbols from simulations of Equations
    \eqref{eq:r_dot} and \eqref{eq:dottheta} with vanishing noise.}
    \label{fig:period_precession}
\end{figure}
The period $T$ of one cycle can be calculated from Equation \eqref{eq:radvel}. The angle can be eliminated using Equation \eqref{eq:sinzvr} and we find for the period the expression
\begin{equation}
T(z_0)= \frac{2}{v_0} \int_{r_{\rm min}}^{r_{\rm max}} {\rm d}r \frac{1}{\sqrt{1-\left(\frac{X(z_0)}{r}\right)^2\exp\left(\frac{2r}{r_c} \right)}}
\label{eq:period}
\end{equation}
Figure \ref{fig:period_precession} presents as solid line the numerical evaluation of the periods in dependence of the absolute value of the angle $|z_0|$ according to Equation 
\eqref{eq:period}. Simulation results of the deterministic version of Equations \eqref{eq:r_dot} and \eqref{eq:dottheta} are shown with the symbol '+'.  
The value decays monotonously for growing $z_0$ performing smaller excursions starting with the escaping trajectory for $z_0$ until it moves along the 
circular orbit at the stationary center $|z_0|=\pi/2$.

Until now, we considered only the reduced $(r,z)$ dynamics. As our particles are moving in the $(x,y)$ plane, we determine now the value
of the position angle $\beta$ in dependence of the distance $r$. Using the equation for the tangential velocity \eqref{eq:dotbeta} and 
Equation \eqref{eq:sinzvr}, 
we write the differential for the angle $\beta$ in dependence of the radial position $r$ as ${\rm d}\beta/{\rm d}r$ and integrate formally 
with initial distance $r_0$ and angle $\beta_0$:

\begin{equation}
\beta(r)=\beta_0+\int_{r_0}^r \frac{{\rm d}r}{r} \frac{1}{\sqrt{\left(\frac{r}{X(z_0)}\right)^2 \exp\left(-2\frac{r}{r_c}\right)-1}}\,.
\label{eq:dbetadr}
\end{equation}

After one cycle the trajectory has moved through the maximal and the minimal distance $r_{\rm max/min}$, reaches again $r_0$ 
and has accumulated an angular shift $\Delta\beta$ of
\begin{equation}
\Delta \beta(z_0)=-2\pi+2\int_{r_{\rm min}}^{r_{\rm max}} \frac{{\rm d}r}{r} \frac{1}{\sqrt{\left(\frac{r}{X(z_0)}\right)^2 \exp\left(-2\frac{r}{r_c}\right)-1}}.
\label{eq:deltabeta}
\end{equation}
Values of the shift at the extremal elongations in dependence of the initial angle difference $z_0$ are presented in Figure \ref{fig:period_precession}.
The shift according to Equation \eqref{eq:deltabeta} is shown as line, while the symbol corresponds to the 
deterministic case of the system according to Equations 
\eqref{eq:r_dot} and \eqref{eq:dottheta}.

\subsection{Dynamics in the $(x,y)$ plane}
\label{ssec:dynamics_xy}
The motion in the $(x,y)$ space is reminiscent of the apsidal precession of the planetary motion, where the aphelion and perihelion shift 
during one revolution around the center of gravity. We elaborate on this analogy in appendix \ref{app:celes}. 
The shape of the trajectories is reminiscent of Lissajous curves, or Rose curves and the motion is in general quasiperiodic. An 
interesting value is the time $T_{ros}$ needed for the precession to rotate more than $2\pi$ meaning that a rosette has completely formed. 
One can estimate this time after determining the number of leaves $K(z_0)$ necessary for the completion of the rosette. 
It reads $K(z_0)=2\pi/\Delta \beta(z_0)$. Therefore, it becomes
\begin{equation}
T_{ros}(z_0)= \frac{2\pi}{\Delta \beta(z_0)}  T(z_0)\,,
\label{eq:Tros}
\end{equation}
with expressions from \eqref{eq:period} and \eqref{eq:deltabeta}.

 
Finally, we can express the bounded trajectories in the $(x(r),y(r))$ plane through the parameter $r$
\begin{equation}
x(r)=\begin{cases} r_c\cos\left(\beta\right), \hspace{15pt} z_0=\pm\pi/2,\hspace{7pt}r_0=r_c, \hspace{15pt} 
\forall \beta\in[0,2\pi)
      \\r\cos\left(\beta(r)+l\Delta\beta \right),  \hspace{15pt} \text{otherwise}
          \end{cases}
 \label{eq:xofr}         
\end{equation}
\begin{equation}
y(r)=\begin{cases} r_c\sin\left(\beta\right),\hspace{15pt} z_0=\pm\pi/2,\hspace{7pt}r_0=r_c, \hspace{15pt} 
\forall \beta\in[0,2\pi)
      \\r\sin\left(\beta(r)+l\Delta\beta \right),  \hspace{15pt} \text{otherwise}
          \end{cases}
 \label{eq:yofr}
\end{equation}
where $\beta(r)$ is due to \eqref{eq:deltabeta} and  with $r\in[r_{\rm min},r_{\rm max}]$, and the positive number of revolutions around the home 
 $l\in \rm N^+$. The first case is the circular trajectory. If the home $(x_h,y_h)$ is not situated at the 
 origin of the coordinate system one can shift $x'(r)=x(r)-x_h$ and $y'(r)=y(r)-y_h$.  

\begin{figure}[h]
    \includegraphics[width=0.32\linewidth]{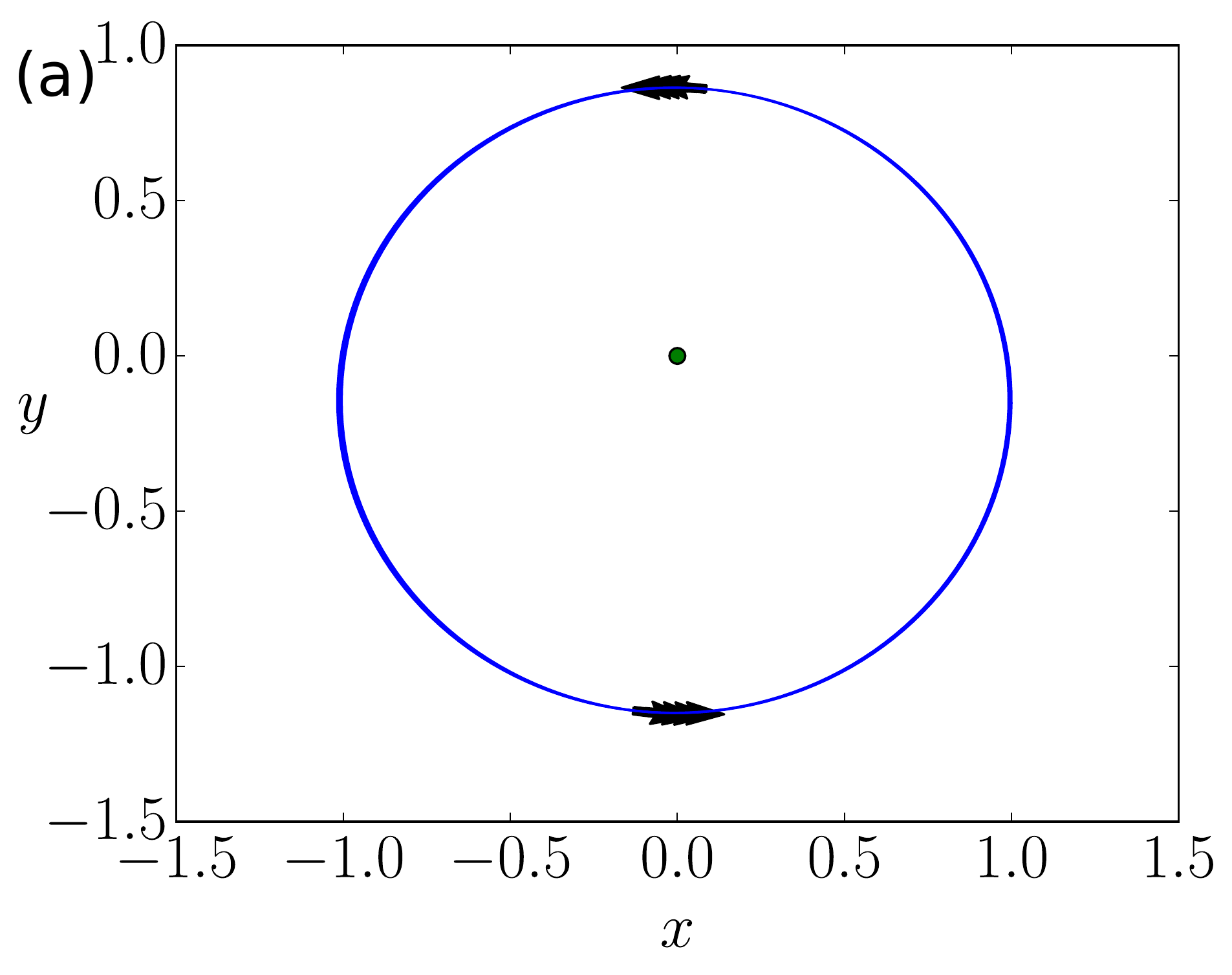}
    \includegraphics[width=0.32\linewidth]{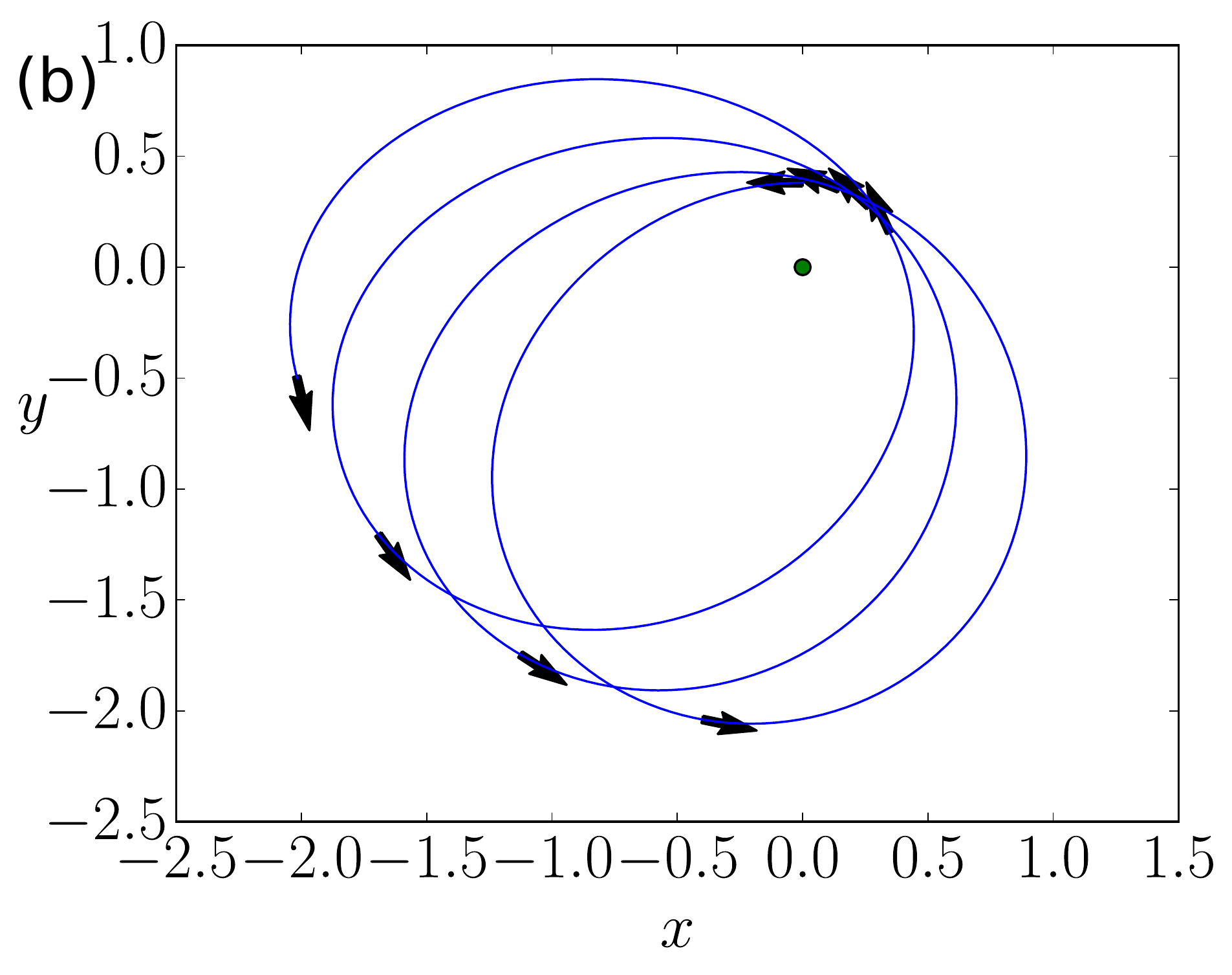}
    \includegraphics[width=0.3\linewidth]{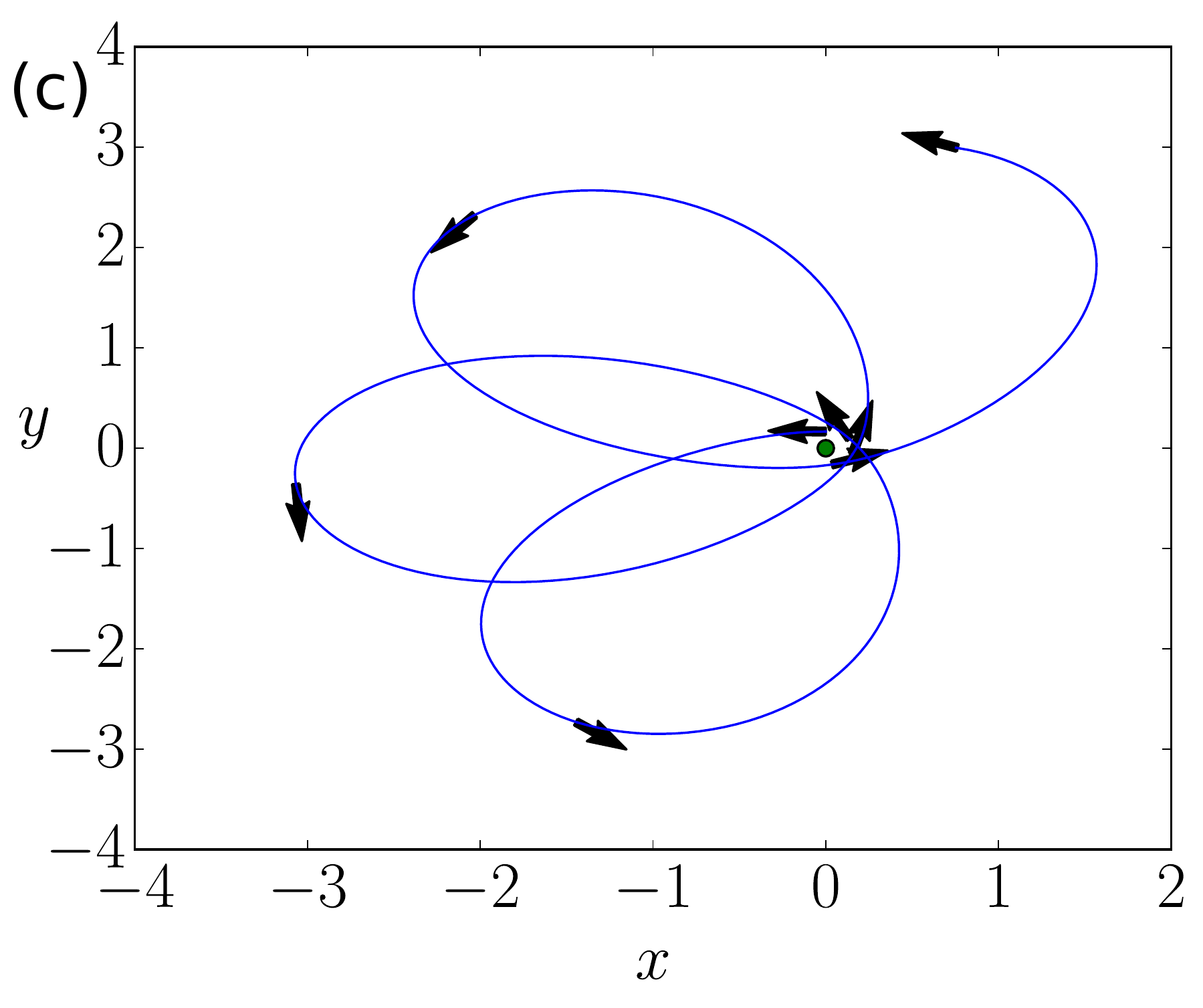}
    \caption{Sample trajectories in $(x,y)$ plane, green dot marks the home. The particles for all trajectories
     start at the minimal distance from the home $x_0=0$, $y_0=r_{\rm min}(z_0)$ and the trajectories are plotted till the maximal distance is reached the fourth time. Shown are three different initial $z_0$.
    Left to Right: $z_0=\pi/2.2$, $z_0=\pi/4$, $z_0=\pi/8$. The arrows indicate the direction of motion. The inverse motion is also possible.
    Parameters: $v_0=1$, $\kappa=1$, $r_0=1$. }
    \label{fig:x_y_plane}
\end{figure}
Figure \ref{fig:x_y_plane} shows sample trajectories according to Equations \eqref{eq:xofr} and \eqref{eq:yofr}, 
with $r_0=r_c$ and $v_0=1$, $\kappa=1$. The location of the home 
is marked as green dot. 
The pictures correspond to three different initial angles $z_0$, i.e. (a) $z_0=\pi/2.2$, (b)  $z_0=\pi/4$, (c) $z_0=\pi/8$. 
The initial position in the $(x,y)$ plane is 
chosen by $x_0=0$ and $y_0=r_{\rm min}(t_0)$. The arrows indicate the direction of motion for these specific initial conditions.
While for $z_0=\pi/2.2$ the trajectory is almost circular, it becomes more and more stretched the smaller $z_0$ becomes, 
the maximal distance increases, 
while the minimal distance gets closer to zero.

\section{The stochastic model}
\label{sec:model_noise} We will now return to the system
  with noise. The equations of motion are given \eqref{eq:dotr},\eqref{eq:dotbeta1} and \eqref{eq:z_stoch}.
    Compared to the deterministic case, one has now two more parameters,
  (i) the parameter $\alpha$ for the noise type and (ii) the parameter
  $\sigma$ - the noise strength.  It will be of importance that the
  noise acts only on the $\theta(t)$ dynamics, respective $z(t)$ dynamics, of the motion 
  and that the radial dynamics is
  perpendicular to the $z$ dynamics. The speed of the particle
  is always constant.

At first one finds that an unbounded motion becomes unlikely in the
  stochastic system. Trajectories remain with high probability at
  finite distances from the home. Therefore, generally the noise
  stabilizes the motion of the searcher.

\subsection{Stochastic dynamics} 

\begin{figure}[h]
    \includegraphics[width=0.4\linewidth]{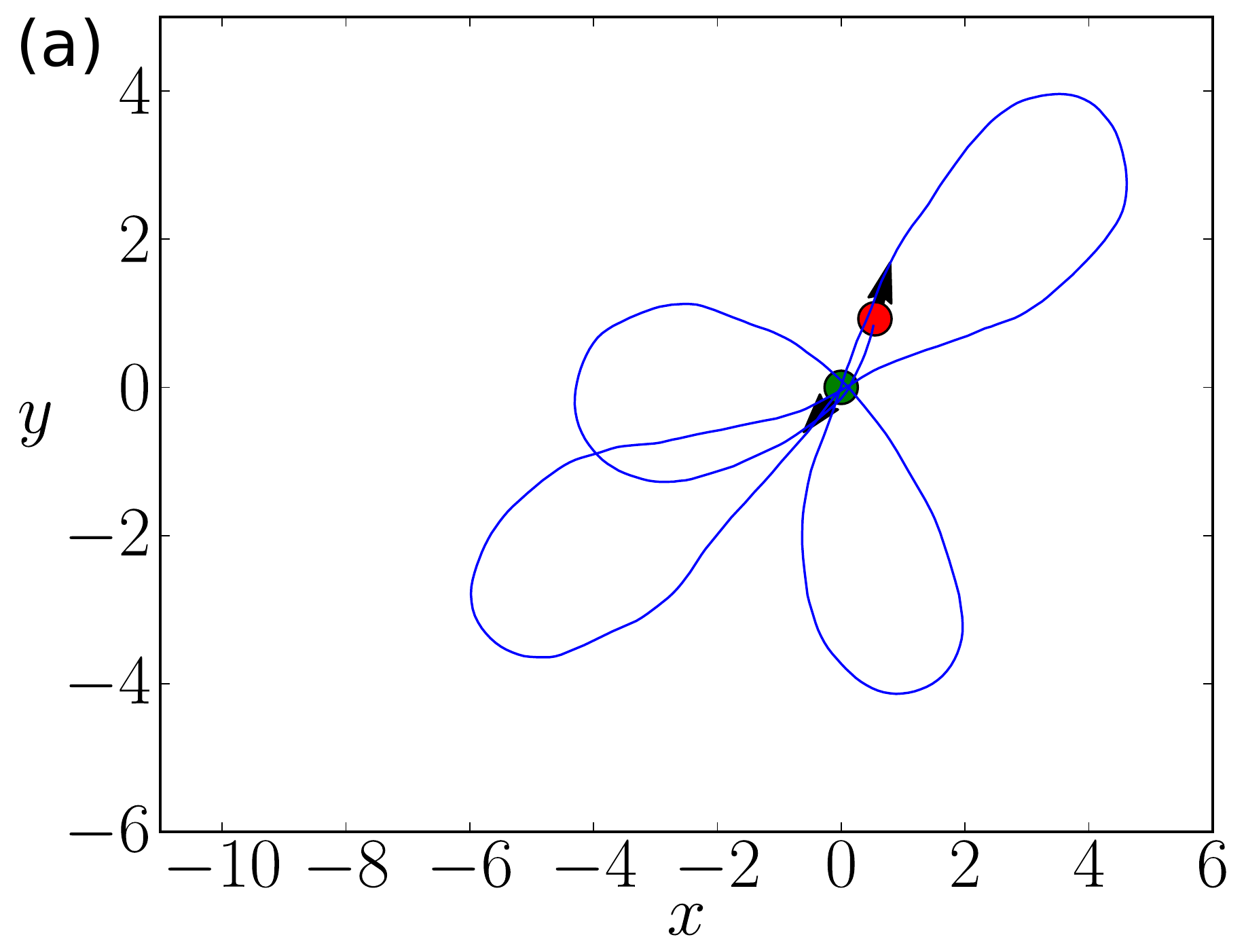}
    \includegraphics[width=0.4\linewidth]{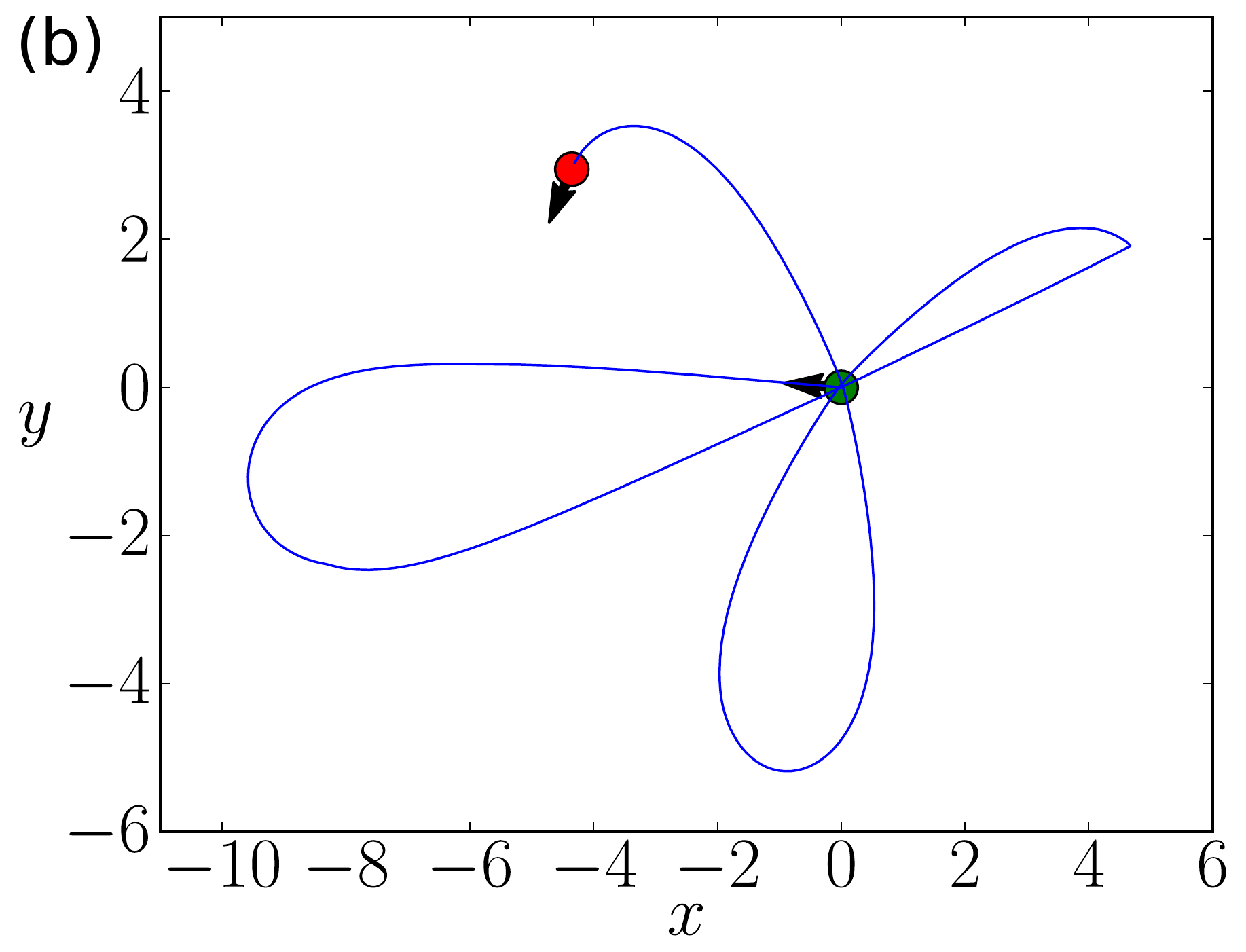}
    \caption{Sample trajectories with small noise strength
      $\sigma^\alpha=0.01$. 
      Trajectory in the $(x,y)$ plane. Green dot marks
      the home.  Particle starts at $(0,0)$ and moves for $\Delta
      t=50$ (red dot). (a) $\alpha=2$, (b) $\alpha=0.5$. 
        Other Parameters: $v_0=1$, $\kappa=1$.
    }
    \label{fig:plane_lownoise}
\end{figure}
\begin{figure}[h]
    \includegraphics[width=0.4\linewidth]{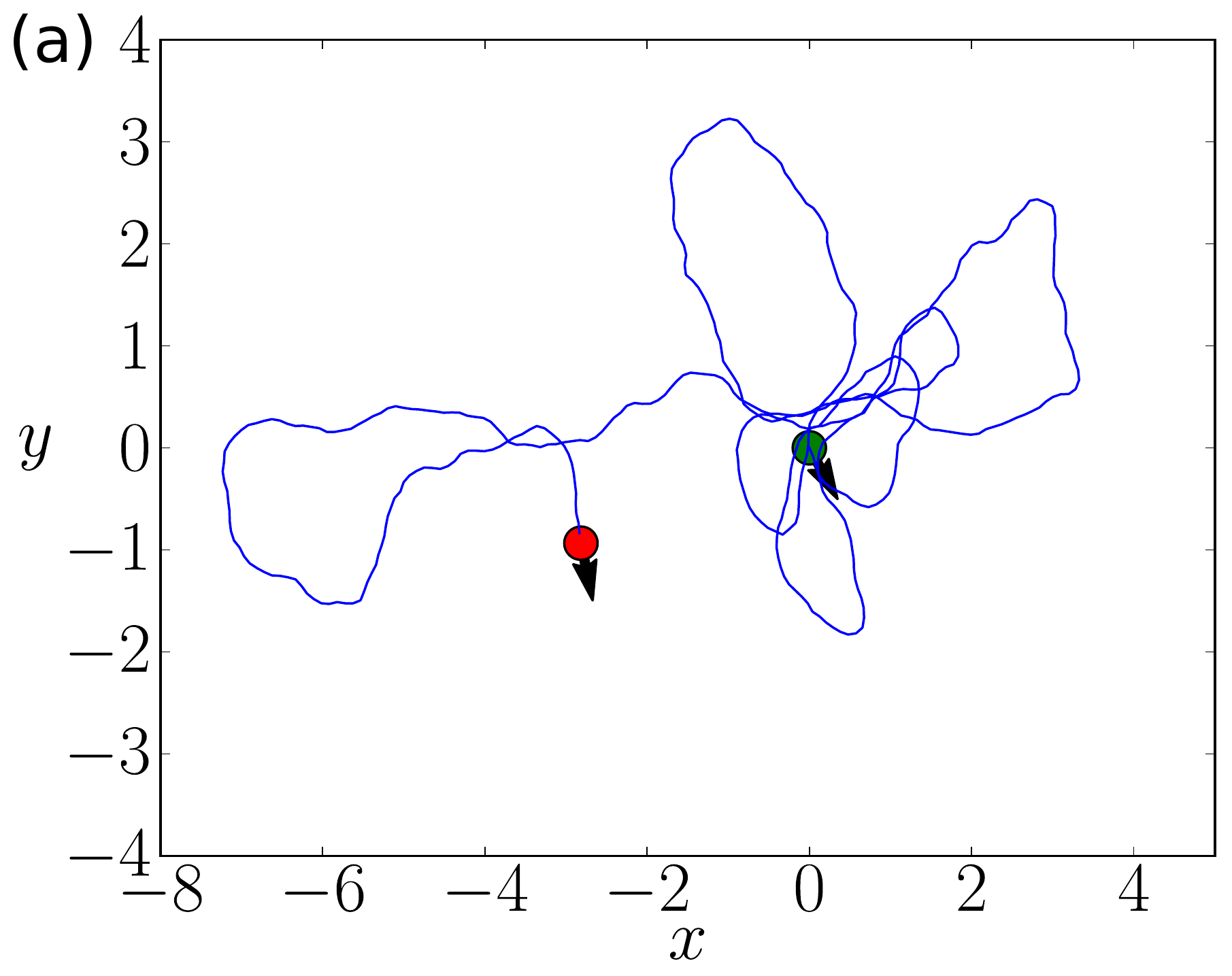}
    \includegraphics[width=0.4\linewidth]{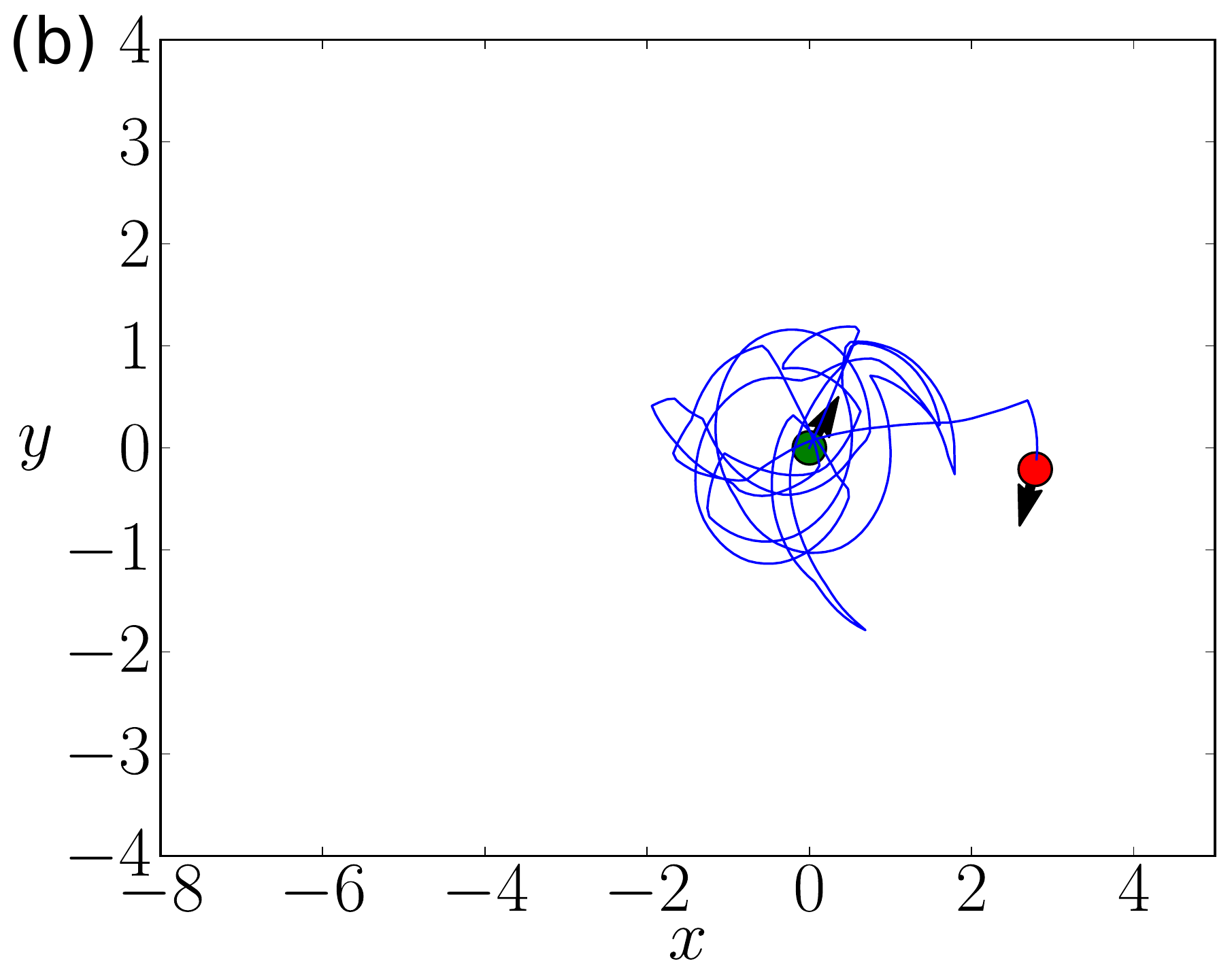}
    \caption{Sample trajectories with large noise $\sigma^\alpha=0.5$ in $(x,y)$ plane. Particle starts at $(0,0)$ (green dot) and moves for $\Delta t=50$ (red dot). 
    (a) $\alpha=2$, (b) $\alpha=0.5$, other parameters: $v_0=1$, $\kappa=1$.   }
    \label{fig:x_y_plane_highnoise}
\end{figure}
Only the noise
type controls jumps of the heading angle. This behavior is
illustrated in Figure \ref{fig:plane_lownoise}.  
We show sample trajectories in the $(x,y)$ plane for two
different noise sources: Gaussian white noise $\alpha=2.0$ on the left and $\alpha=0.5$ on the right. The noise strength is chosen such that the
influence of the deterministic part of the time evolution of the
heading \eqref{eq:dottheta} is still recognizable,
$\sigma^\alpha=0.01$. The particle starts at the home, marked as green
dot at $(x,y)=(0,0)$ and a time frame of $\Delta t=50$ is shown. 
While the trajectory on the left is wiggling, due to a lot of small noise induced changes in
the direction of motion, the trajectory 
on the right with small $\alpha=0.5$ is rather smooth
corresponding to almost no change in the deterministic heading and
with one clear sudden jump at $(x,y)=(4,2)$. 

Increasing the noise strength, suppresses the deterministic influence on the trajectories, as can be seen in Figure \ref{fig:x_y_plane_highnoise}.   
The noise strength $\sigma^\alpha=0.5$ was changed compared with the previous Figure \ref{fig:plane_lownoise}. 
The particles again started at the home $(x,y)=(0,0)$ and moved for the same time interval $\Delta t= 50$. The deterministic part of the motion is 
no longer visible and also the trajectory on the right ($\alpha=0.5$) appears to be confined to a smaller region of space. While the trajectory seems 
to be confined to a small region, we show in the following Section \ref{sec:spat_dist}, that the overall probability density function to find a 
particle at a specific point $(x,y)$ is independent of the noise type $\alpha$ and the noise strength $\sigma$.

\subsection{Spatial distribution}
\label{sec:spat_dist}
One measure to characterize the spatial extension of the search is the marginal density of the distance from the home. 
Here we will look for its asymptotic stationary expression as the result of solving of the corresponding Fokker-Planck equation (FPE). 
To find this marginal density we inspect the the probability density function (pdf) of the stochastic dynamics under consideration. 
It is the transition pdf $P(r,z,\beta,t|r_0,z_0,\beta_0,t_0)$ that determines the density in three dimensional space $r,z,\beta$ at 
time $t$, if started with the initial conditions $r_0,z_0,\beta_0$ at $t_0$. Likewise in the deterministic case, we immediately 
realize that the the stochastic $\beta$ dynamics separates from the two other variables $(r,z)$. 
It holds $P(r,z,\beta,t)= P(r,z,t)P(\beta,t|r,z)$ where we have omitted the initial conditions for simplicity, 
meaning that the $r,z$ dynamics is independent of the $\beta$ dynamics. Therefore, to get the marginal spatial density, 
we may restrict to the consideration of  the pdf $P(r,z,t|r_0,z_0,t_0)$ for finding a particle at distance $r$ and having the angle 
$z$ at time $t$ if started at time $t_0$ at distance  $r_0$ with $z_0$.

Being interested in the asymptotic stationary limit when initial conditions are forgotten, 
we omit here and further on the initial conditions and in the notation of the pdf. 
Given Equations \eqref{eq:dotr} and \eqref{eq:dotz}, we can write down the 
corresponding Fokker Planck equation\cite{Ditlevsen,Schertzer}.
\begin{equation}
\frac{\partial}{\partial t}P(r,z,t)=\left[-v_0\frac{\partial}{\partial r}\cos(z)+\frac{\partial}{\partial z}\left(\frac{v_0}{r}-\kappa \right)\sin(z)+\left(\frac{\sigma}{v_0}\right)^\alpha \frac{\partial^\alpha}{\partial |z|^\alpha}\right]P(r,z,t)
\label{eq:fpe_full}
\end{equation}
We assume a steady state $P(r,z,t\to \infty)=P_0(r,z)$, with $\partial P_0(r,z,t)/\partial t=0$ and make a separation Ansatz 

\begin{equation}
P_0(r,z)=P_0(r|z)P_0(z).
\end{equation}
Since the noise spreads the probability homogeneously around the
angular dynamics and as no effective force repels the noisy shifts, no
direction $z$ is preferred. Therefore, we assume that $z(t)$ becomes
equidistributed after the relaxation time $\tau$ from Equation
\eqref{eq:tau} and we set $P_0(z)=1/2 \pi$. Further on, it
turns out that this homogenization in the angular dynamics results
also in an independence of the asymptotic spatial distribution on $z$,
i.e. $P_0(r|z) \longrightarrow P_0(r)$.
Afterwards, we find for the latter radial pdf the equation:
\begin{equation}
0=\left[-\frac{\partial}{\partial r}+\frac{1}{r}-\frac{1}{r_c} \right]v_0\cos(z)P_0(r)\,,
\end{equation}
with $r_c$ from Equation \eqref{eq:r_c}.
The pdf does no longer depend on the angle $z$, so we can drop the cosine function, integrate 
and the radial pdf is given by
\begin{equation}
P_0(r)=\frac{r}{r_c^2}\exp\left(-\frac{r}{r_c} \right).
\label{eq:p0}
\end{equation}
\begin{figure}[h]
    \includegraphics[width=0.45\linewidth]{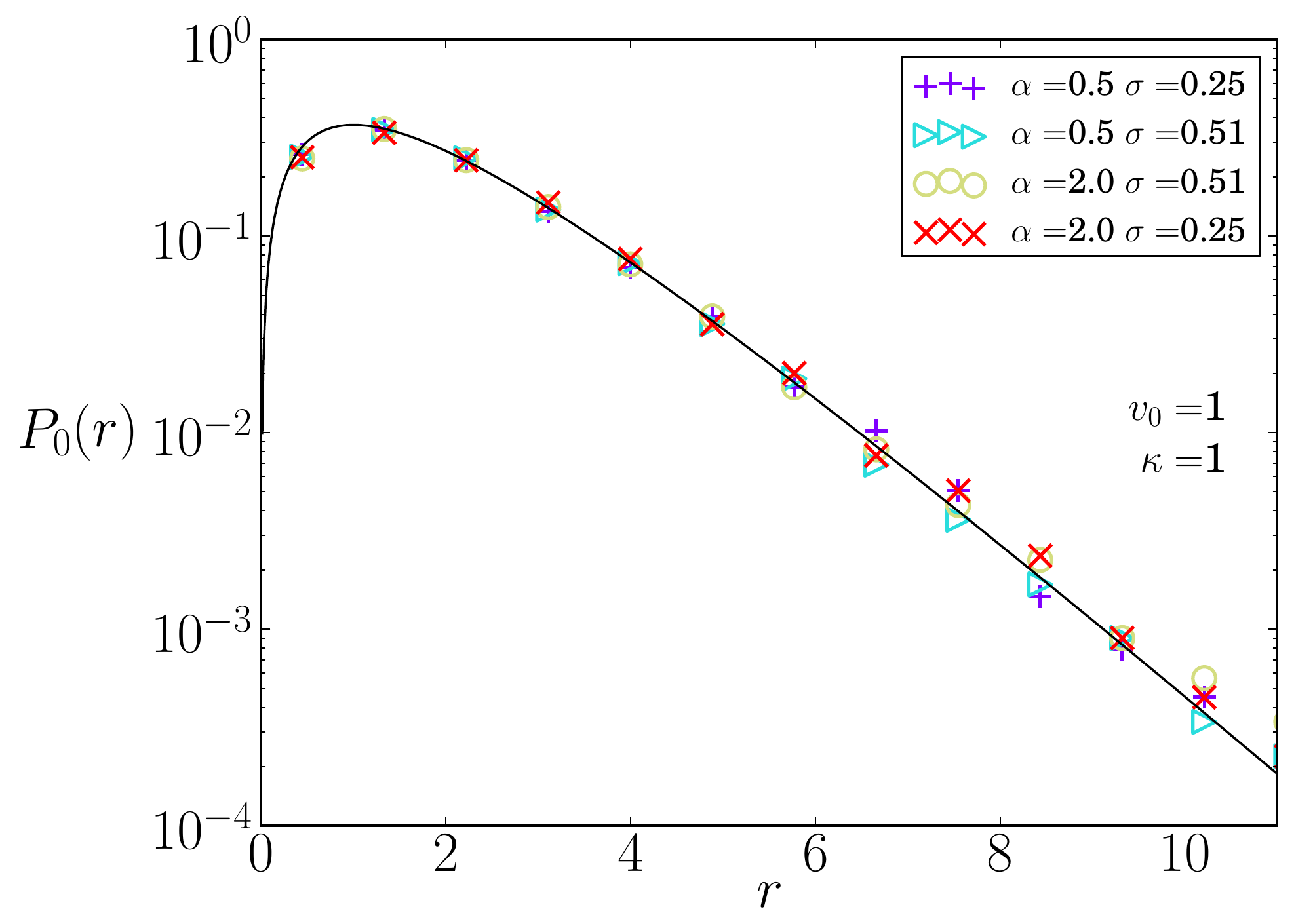}

    \caption{The stationary radial pdf  $P_0(r)$. Symbols from simulations of Equations \eqref{eq:r_dot} and \eqref{eq:dottheta}, with 
    \eqref{eq:beta}.
    Black line according to Equation \eqref{eq:p0}}
    \label{fig:p0}
\end{figure}
Including the angle $z$, we get
\begin{equation}
P_0(r,z)=\frac{r}{2\pi r_c^2}\exp\left(-\frac{r}{r_c} \right).
\label{eq:p0z}
\end{equation}
Surprisingly, the spatial distribution and therefore the probability to
find a particle at a specific distance from the home, does neither
depend on the noise type, nor on the noise strength. The reason behind
this is that the noise acts perpendicular to the motion of particles.
The distribution \eqref{eq:p0} is even valid and holds true for
$\sigma=0$ if the system is initially distributed according to
\eqref{eq:p0}.
The spatial distribution $P_0$ has a maxima at $r=r_c=v_0/\kappa$. Increasing the coupling strength $\kappa$ and keeping the speed $v_0$ 
fixed shifts the maxima closer to the home and 
the peak becomes more pronounced. The stronger the coupling towards the home the closer to the home is the maxima situated. 
Increasing the speed $v_0$  leads to a growth of the distance between home and maxima and it broadens 
the spatial distribution $P_0(r)$.

Returning to Cartesian coordinates the pdf reads:
\begin{equation}
P_0(x,y)=\frac{1}{2\pi r_c^2}\exp\left(-\frac{1}{r_c}\sqrt{x^2+y^2} \right)
\end{equation}
This pdf has maximal probability density to find the particle
at the home $(0,0)$. We mention that the stationary pdf of the angle $\beta$ is also uniform at all distances.

Figure \ref{fig:p0} shows the stationary radial pdf $P_0$ from \eqref{eq:p0} as line in comparison with simulation results of Equations
\eqref{eq:r_dot} and \eqref{eq:dottheta}, with the position angle defined as \eqref{eq:beta}. The simulations confirm that indeed the 
radial pdf is independent of noise type $\alpha$ and noise strength $\sigma$. We underline that the stationary spatial density qualitatively agrees with the experimental findings for the observed residence 
probability of a fruit fly as reported in \cite{Kim_Dickinson_2017}. Taking the speed $v_0=10mm/s$ from \cite{Kim_Dickinson_2017} and judging from Figure S4 in \cite{Kim_Dickinson_2017}, we consider $\kappa=0.5/s$ to be realistic values for a fly.

\subsection{Relaxation time $\tau$}
\label{sec:tau}
\begin{figure}[h]
    \includegraphics[width=0.45\linewidth]{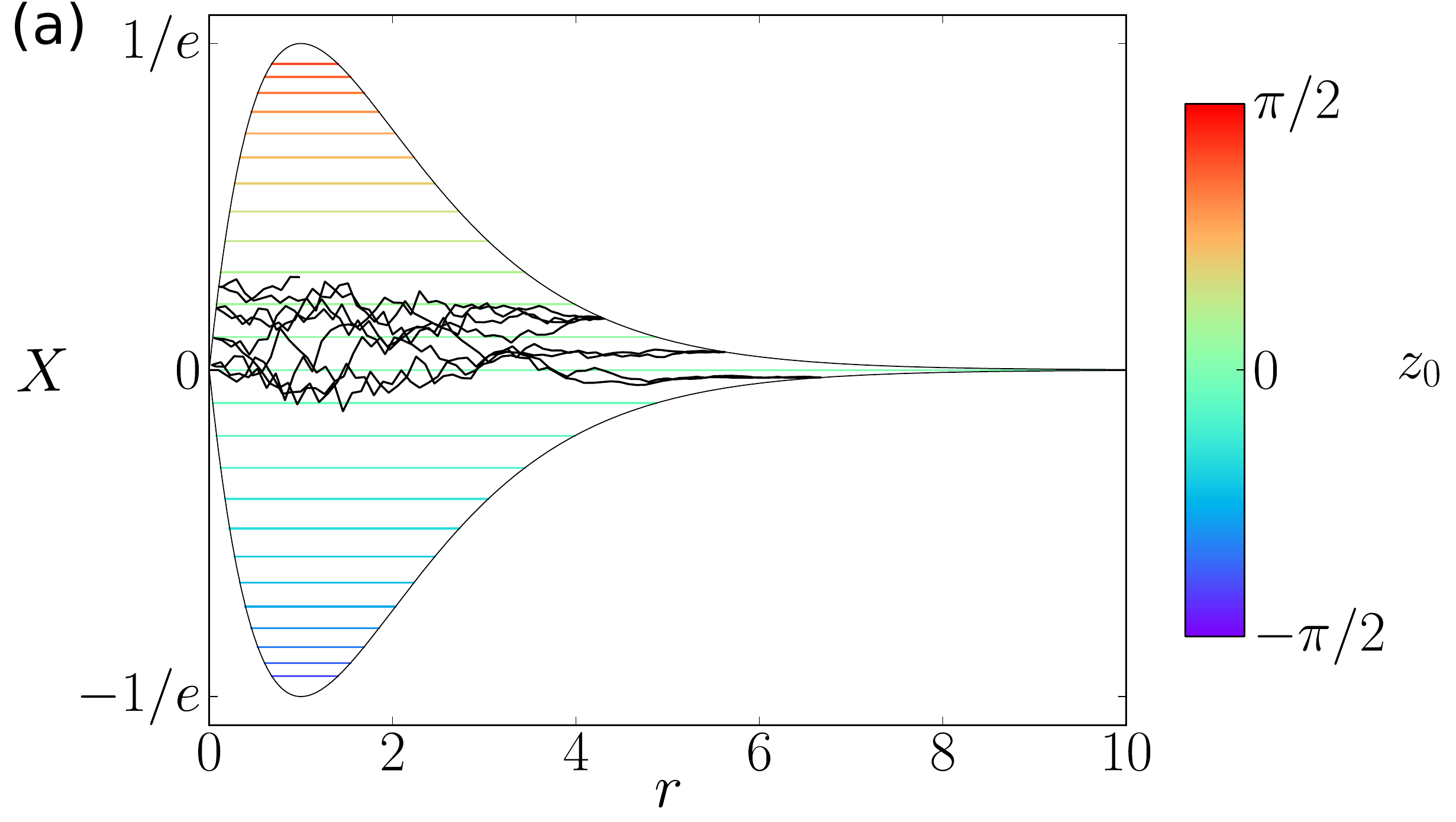}
     \includegraphics[width=0.45\linewidth]{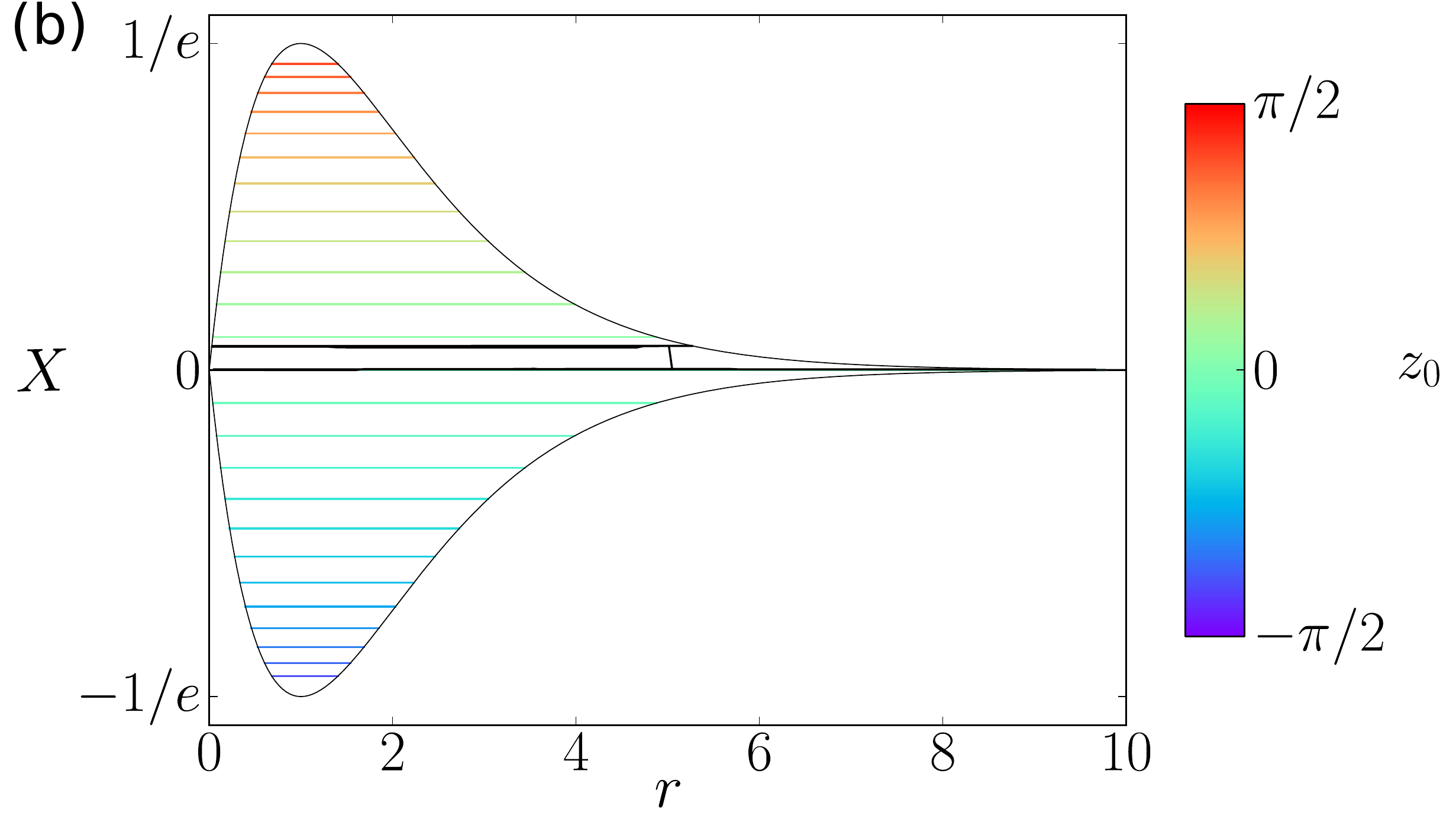}
    \includegraphics[width=0.45\linewidth]{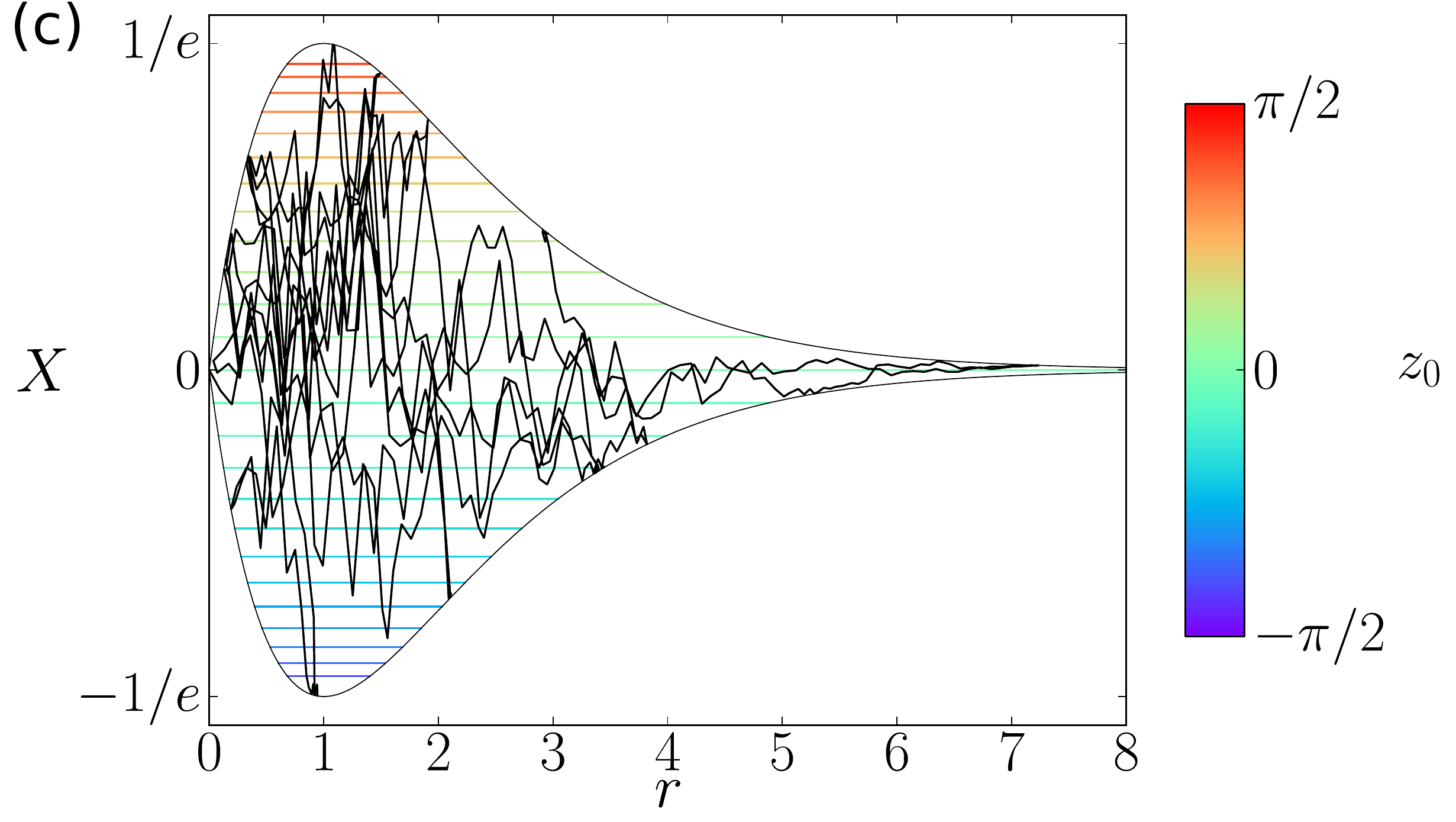}
     \includegraphics[width=0.45\linewidth]{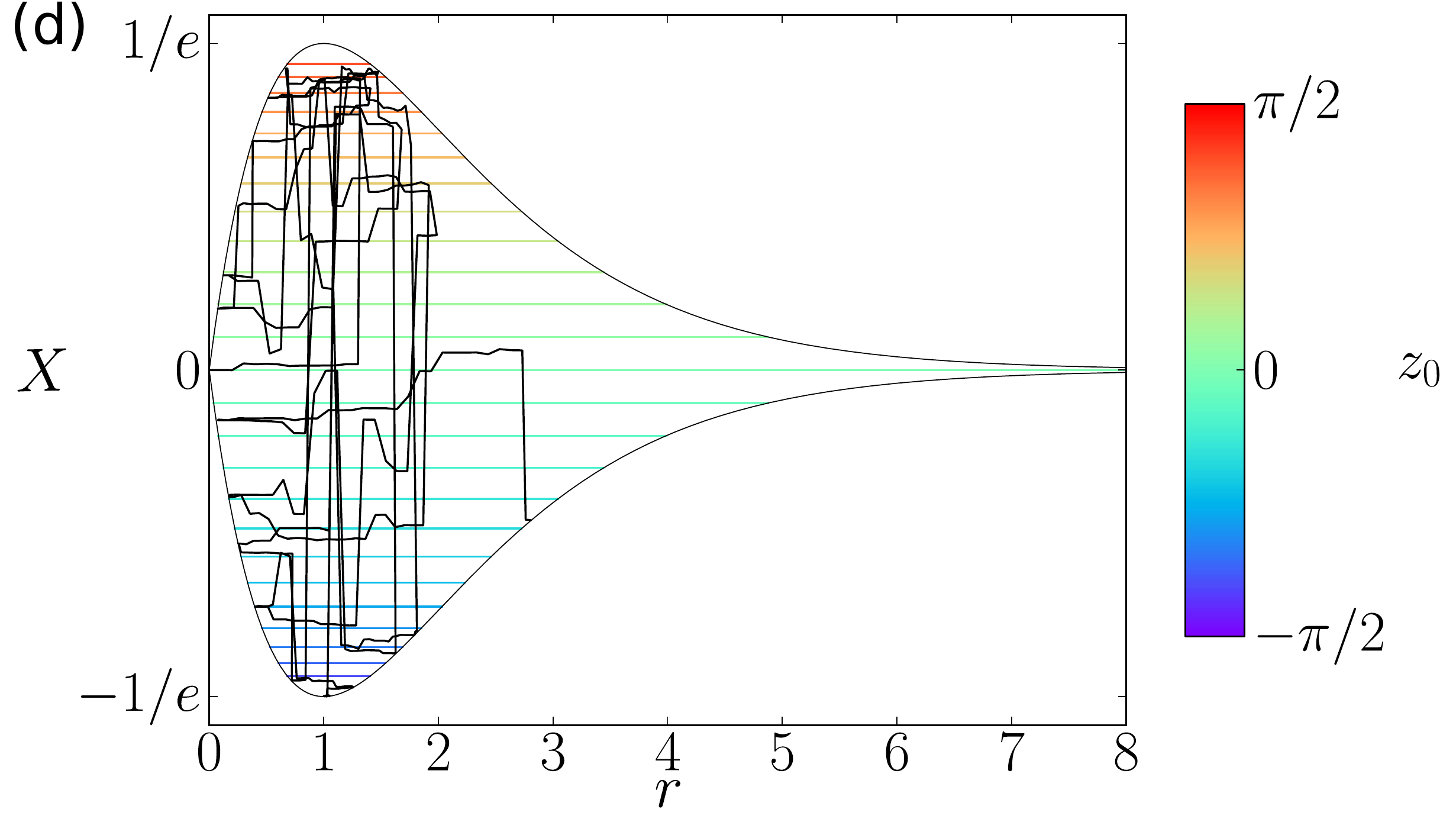}
    \includegraphics[width=0.45\linewidth]{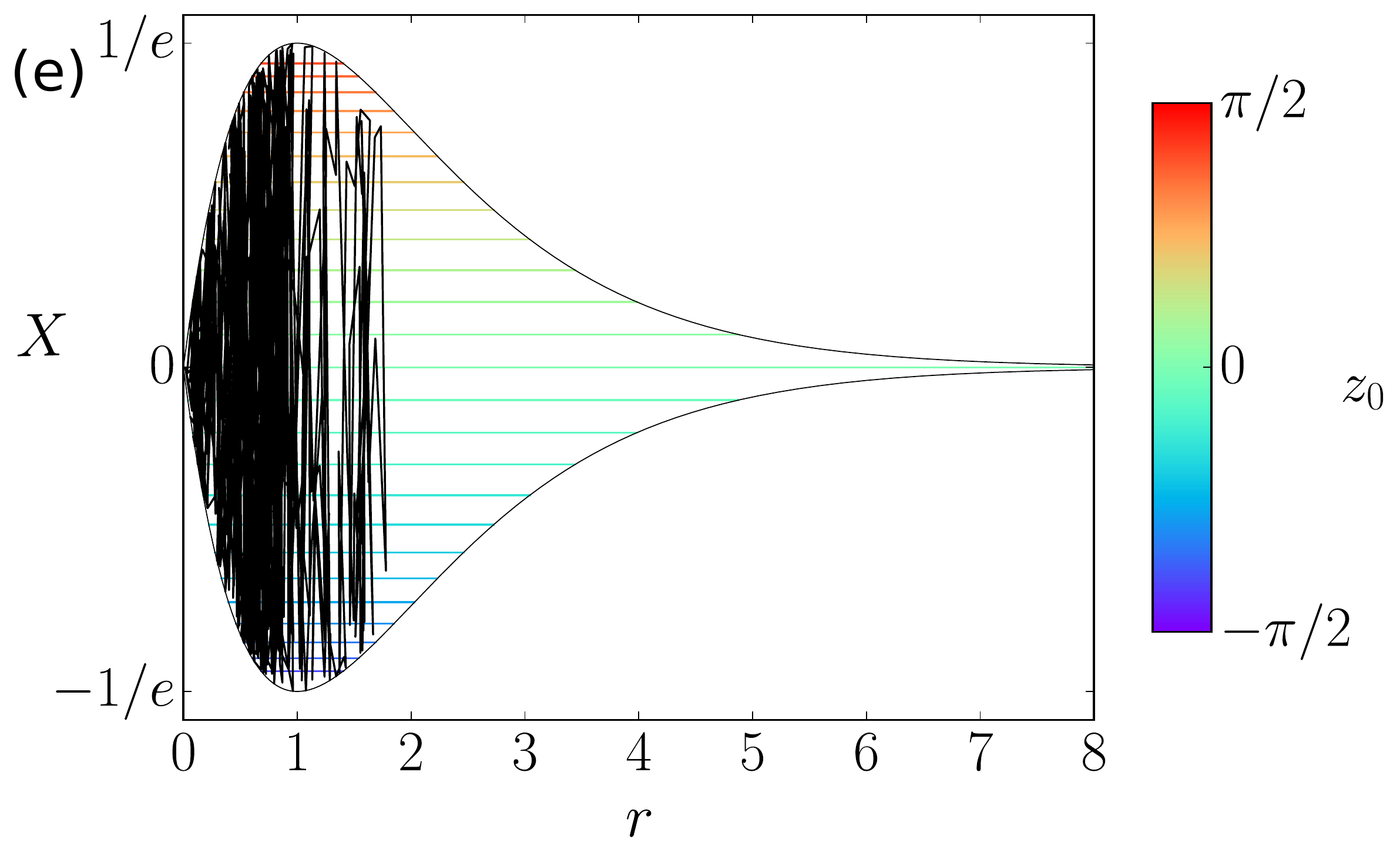}
     \includegraphics[width=0.45\linewidth]{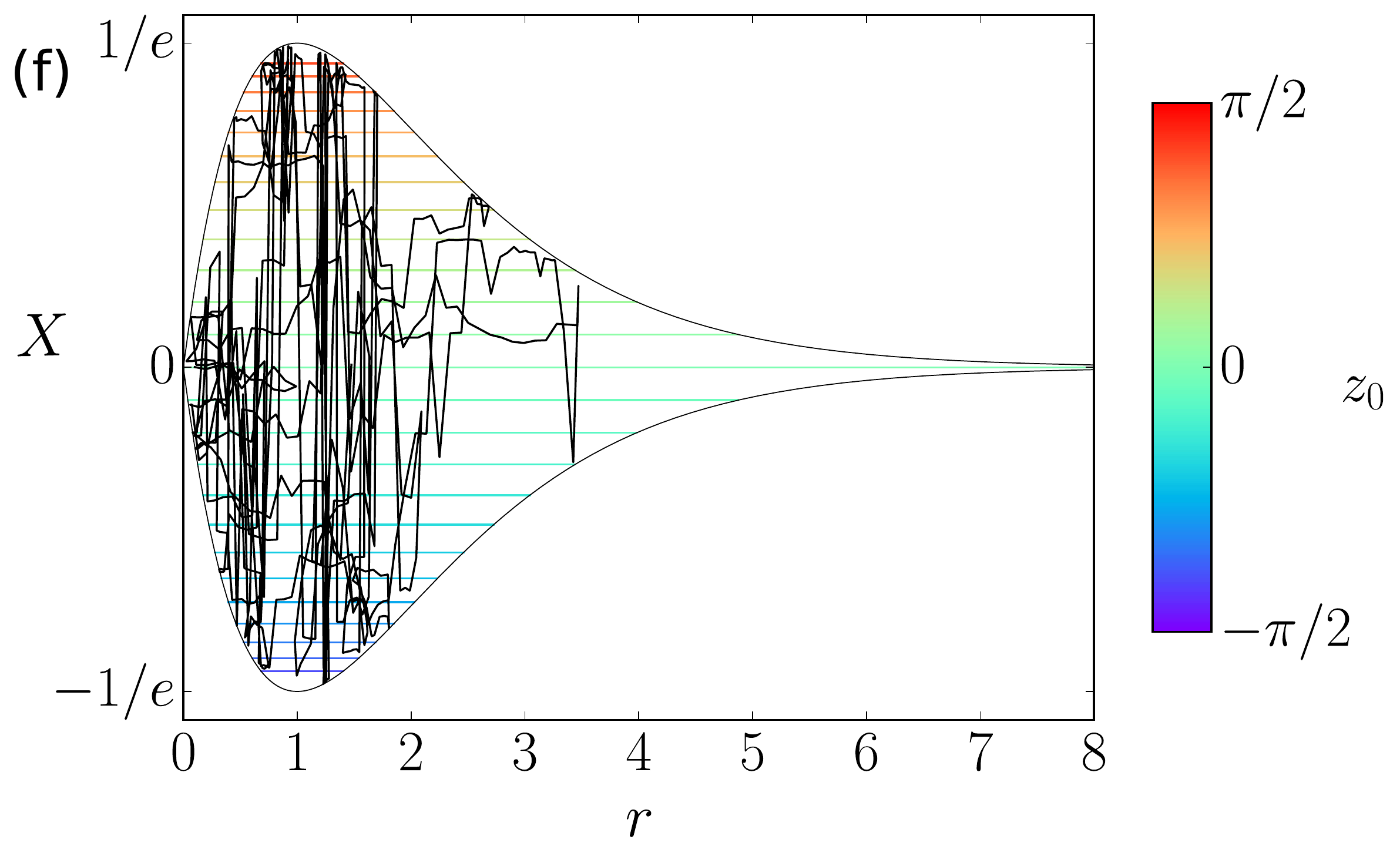}
    \caption{Top: Trajectories of Figure \ref{fig:plane_lownoise} with $\sigma^\alpha=0.01$ in $(r,X)$ plane. 
    Middle: Trajectories of Figure \ref{fig:x_y_plane_highnoise} with $\sigma^\alpha=0.5$ in $(r,X)$. Bottom: Sample trajectories with high noise strength $\sigma^\alpha=4.0$.
    Total trajectory 
    length is $t=50$ corresponding to $t=0.5 \tau$ (Top) and $t=25\tau$ (Middle) and 
    $t=200\tau$ (Bottom), with $\tau$ being the relaxation time from \eqref{eq:tau}. 
    Color coded straight horizontal lines are various deterministic paths with constant $X(z_0,r_0=r_c)$ values. Colors correspond to the deterministic motion
    in the $(r,z)$ plane of Figure \ref{fig:r_z_space}. Left: $\alpha=2$, Right: $\alpha=0.5$, 
    Other parameters: $v_0=1, \kappa=1$.}
    \label{fig:r_X_noise}
\end{figure}

In order to investigate how the stationary radial pdf is approached, 
it is useful to consider the integral of motion $X(z_0,r_0)$ from \eqref{eq:sinzvr} together with the distance $r$ as variables, as shown in 
Figure \ref{fig:r_X_noise}. Deterministic trajectories with initial conditions $(r(t=0)=r_0, z(t=0)=z_0)$ 
become straight lines in the $(r,X)$ space, as $X$ is a constant,
i.e. $\dot{X}=0$.
Deterministic trajectories with different initial conditions $z_0$ and $r_0=r_c$ are shown in the $(r,X)$ space in Figure \ref{fig:r_X_noise} as  
color coded straight lines. The colors correspond to the respective trajectories of Figure \ref{fig:r_z_space}.

The envelope equals the not normalized spatial distribution \eqref{eq:p0}. The transformation from the $(r,z)$ plane to the $(r,X)$ plane 
does not preserve the direction of motion towards and away from the home, both happen on the same line.

In the noise driven system the value of $X$ looses the meaning of an integral of motion. It becomes stochastic and, in consequence,  
time dependent. The noise allows for a vertical motion in Figure \ref{fig:r_X_noise}. 
The black trajectories in Figure \ref{fig:r_X_noise} show the previous trajectories of 
Figure \ref{fig:plane_lownoise} with low noise intensity $\sigma^\alpha=0.01$ in the top row and
at in the middle row the trajectories of Figure \ref{fig:x_y_plane_highnoise} with $\sigma^\alpha=0.5$. 
In the bottom row, we show as black line sample trajectories for $\sigma^\alpha=4.0$.
On the left the noise type corresponds to Gaussian white noise $\alpha=2$ and on the right to $\alpha=0.5$. 

Three things should be noted, when looking at the pictures: (I) The different noise types act differently, while Gaussian white noise 
causes rather continuous small vertical changes, the noise with $\alpha=0.5$ acts mostly through sudden jumps, as can be clearly seen
in the top and middle row. The trajectories for $\alpha=0.5$ stay a rather long time with a specific 
deterministic trajectory compared to the Gaussian noise case. 
(II) The noise strength influences the time needed for a vertical motion of the particle, as can 
be seen following the pictures in a column. With increasing noise strength the particles distribute faster. (III) In radial 
direction however an increasing noise strength slows down the motion of the particle, which is particular clearly visible in the 
Gaussian white noise column.

Interestingly, the relaxation time $\tau$ is equal for the two trajectories in each row, as we will show next, 
although their behavior appears quite different. 

\begin{figure}[h]
    \includegraphics[width=0.45\linewidth]{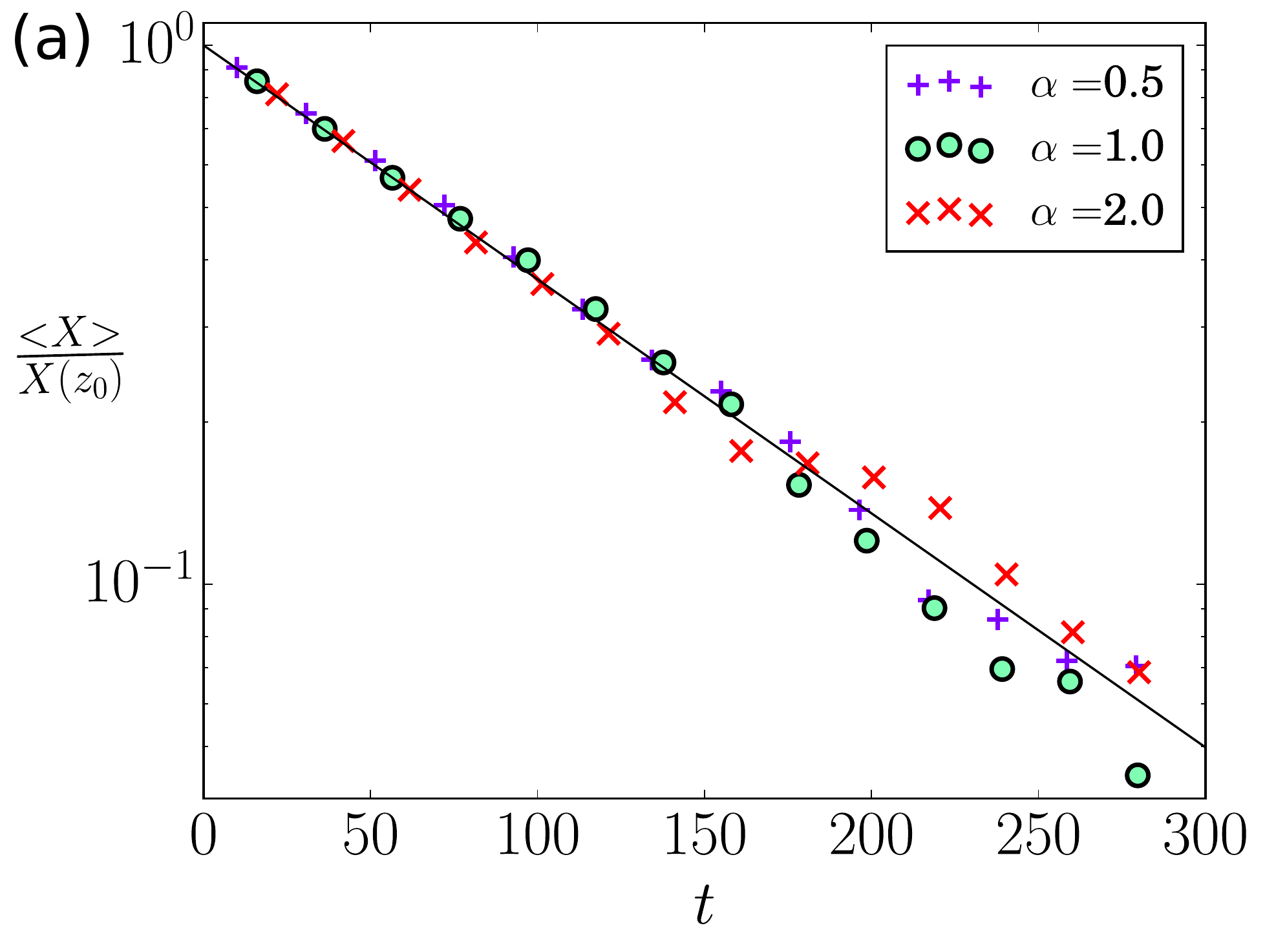}
    \includegraphics[width=0.45\linewidth]{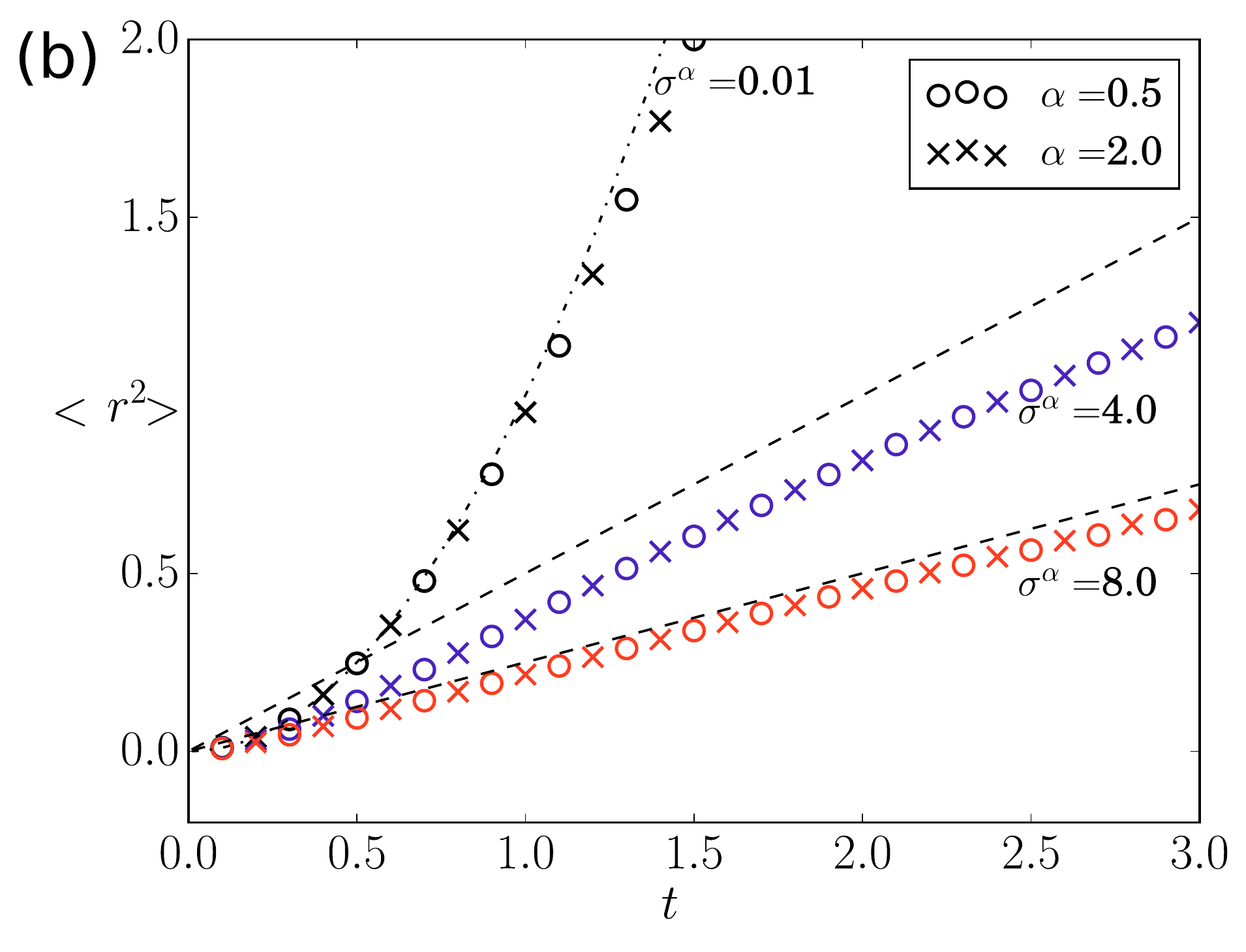}
    \caption{(a) Symbols show the normalized ensemble average of $X$ obtained from simulation of \eqref{eq:r_dot} and \eqref{eq:dottheta}, 
    with $v_0=1$, $\kappa=1$ and $\sigma^\alpha=0.01$ $(\tau=100)$ and initial $X(t=0)=X(z_0)$.
    Black line: exponential decay according to \eqref{eq:avXt} (b) Mean squared displacement $<r^2>$ from simulations of particles started at $(x_0,y_0)=(r_c,1)$
    as at this distance the deterministic influence vanishes, with $r_c=1$, as $v_0=1$, $\kappa=1$ for two different values of $\alpha$, each drawn for three different values of $\sigma^\alpha$. Dashed
    line corresponds to Equation \eqref{eq:msd}. The dotted dashed line corresponds to a ballistic grows $\propto t^2$. The plotted time frame 
    is well below the steady state spatial distribution is reached.}
    \label{fig:tau}
\end{figure}

The vertical motion, i.e. the motion in $X$ direction, can be characterized by the time evolution of the normalized ensemble average 
\begin{equation}
<X>(t)= \int_0^\infty \text{d} r\int_{-\pi}^{\pi} \text{d} z\, X(r,z)\,P(r,z,t|r_0,z_0,t_0)\,
\label{eq:aveX}
\end{equation}
conditioned to the initial values $r_0$ and $z_0$  at $t_0$ with the respective value $X_0=X(r_0,z_0)$ according to \eqref{eq:sinzvr}. 
This evolution is governed by the linear differential equation:
\begin{equation}
\frac{{\rm{d}}}{{\rm{d}} t}<X>=-\frac{1}{\tau}<X>\,.
\label{eq:meanX}
\end{equation}
A derivation of this equation based on the FPE multiplied by $X$ and integrated over $r$ and $z$ is given in the Appendix \ref{app:X-dyn}. 
As result of \eqref{eq:meanX} we obtain:
\begin{equation}
<X(t)>=X_0 \exp\left(-\frac{t}{\tau}\right)\,,
\label{eq:avXt}
\end{equation}
with \begin{equation}
\tau=\left(\frac{v_0}{\sigma}\right)^\alpha \,.
\label{eq:tau}
\end{equation}
This time scale $\tau$ has the meaning of a relaxation time and is originated by the noise, without noise no relaxation takes 
place since  $\tau \propto 1/\sigma^{\alpha} \rightarrow \infty$. With noise any initial state $X_0$ is forgotten for $t\gg\tau$, as can 
be seen in Figure \ref{fig:tau} (a). For three different values of $\alpha$ but with the same relaxation time 
$\tau$ the time dependent averages $<X(t)>$ obtained from simulation of 
the Equations \eqref{eq:r_dot} and \eqref{eq:dottheta} with \eqref{eq:beta} are plotted together with Equation \eqref{eq:avXt} as line. 
Hence, it is the  time $\tau$ which is needed to vertically redistribute a trajectory in Figure \ref{fig:r_X_noise}. 
We also remark that $\tau$ is the time after that a freely moving active particle with $\kappa=0$ has forgotten its initial heading 
direction \cite{noetel:2017}.


Horizontal motion, the motion in radial direction, in Figure \ref{fig:r_X_noise} is also governed by 
the noise dependent time scale $\tau$ 
but by  an effect acting oppositely as in case of the $X$. It means
that with increasing $\tau$ the particles spread slower in radial direction, while they spread faster in $X$ direction 
as was discussed above. We elucidate this effect by deriving an
overdamped description of the nonlinear $(r,z)$ dynamics in the Appendix \ref{app:r}. We call it an "overdamped" description as 
the angle $z$ is the fast variable and the dynamics of the distance $r$ is assumed to be slow. 
The equation which determines then for $t\gg\tau$ the evolution of the marginal radial pdf $P(r,t|r_0,t_0)$ is a Smoluchowski equation, 
all dependencies on the various $\alpha$-values of the noise are expressed through the relaxation time $\tau$ as defined in \eqref{eq:tau}. 
It becomes
\begin{eqnarray}
&&\frac{\partial}{\partial t}P= D_{\rm{eff}}\frac{\partial}{\partial r}\left(\frac{\partial}{\partial r}P-\left(\frac{1}{r}-\frac{\kappa}{v_0} \right)P\right)\,,
\label{eq:fpe_over1}
\end{eqnarray}
with the effective diffusion coefficient 
\begin{equation}
D_{\rm{eff}}=\frac{v_0^2 \tau}{2}\,.
\label{eq:deff_search}
\end{equation}
The overdamped approximation is valid if the effective diffusion coefficient remains finite.
The expression \eqref{eq:deff_search} of the latter coincides with results for freely moving active particles $\kappa=0$ with 
angular driving by $\alpha$-
stable noise \cite{noetel:2017}.

Note, that the overdamped description is solved for the steady state by our previous solution 
\eqref{eq:p0}. We also underline, that the spatial relaxation is proportional the relaxation time, especially 
around  $r\approx r_c$ follows the mean squared displacement (MSD) 
\begin{equation}
\langle r^2 \rangle	\sim v_0^2\tau t 
\label{eq:msd}
\end{equation}

Figure \ref{fig:tau} (b) displays as symbols the MSD obtained from simulations for particles started 
at $(x(t=0),y(t=0))=(r_c,0)$ and as dashed line Equation \eqref{eq:msd}, noise strength and type as indicated in the figure. 
The displayed time frame
is small against the time scale for the establishment of the steady state.
With increasing noise strength the simulation results for MSD align with \eqref{eq:msd}. This implies that with increasing noise strength $\sigma$ the relaxation in 
radial direction slows down. We also show the MSD for a small noise strength $\sigma^\alpha=.01$. There the particle moves in the 
ballistic regime and the MSD grows $\propto v_0^2 t^2$, as indicated by the dot dashed line. 
The different dependence on the relaxation time of the radial and vertical motion in the $r,X$ plane will be the key element for the existence
of an optimal time for local search.

\section{Local search }
\label{sec:disc}
Local search 
consists of a search and a return part. In our model both tasks are described by the same stochastic dynamics. 
We discuss now the return
and, afterwards, the search part. 

\subsection{Sojourn time}
\begin{figure}[h]
    \includegraphics[width=0.47\linewidth]{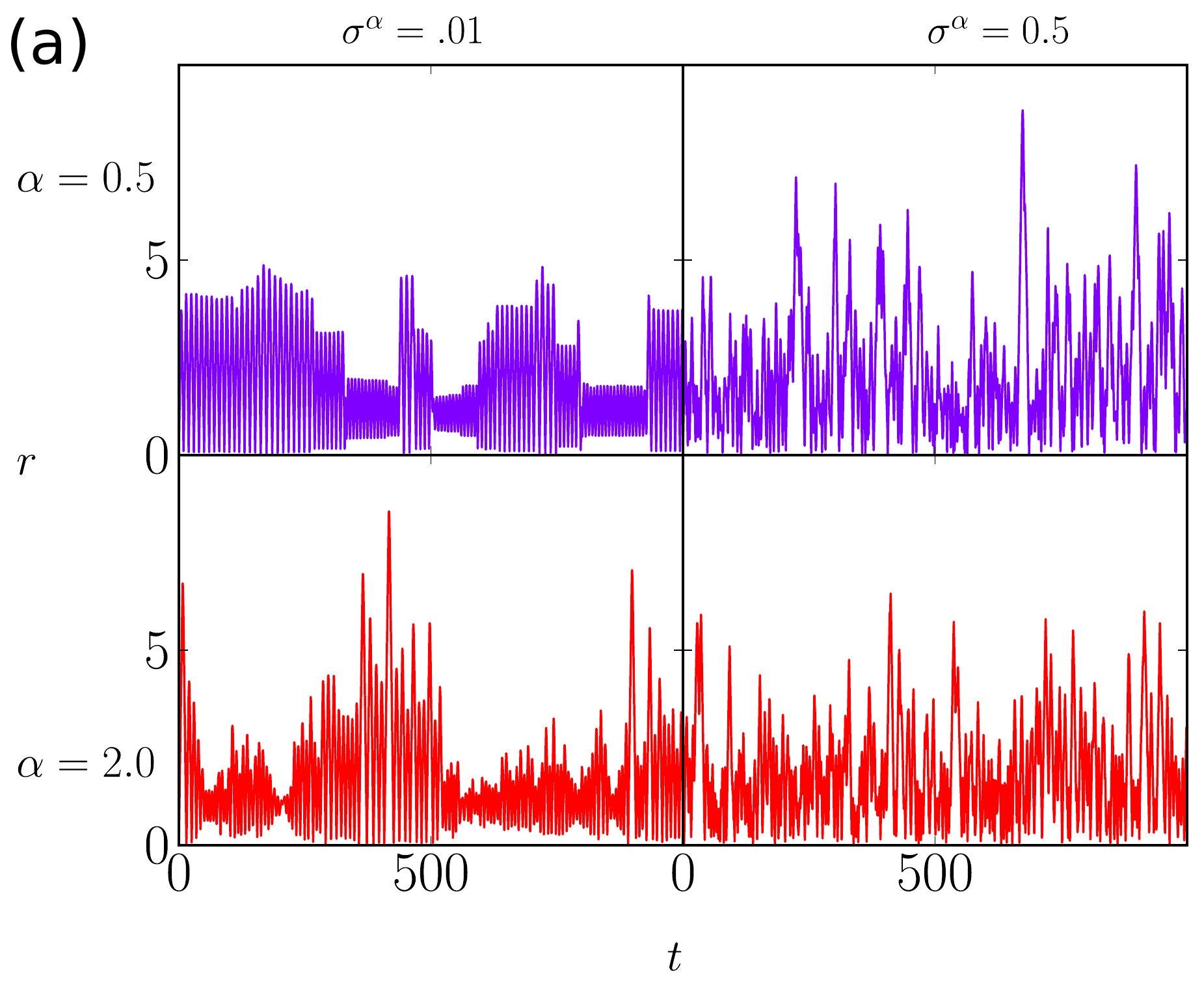}
    \includegraphics[width=0.49\linewidth]{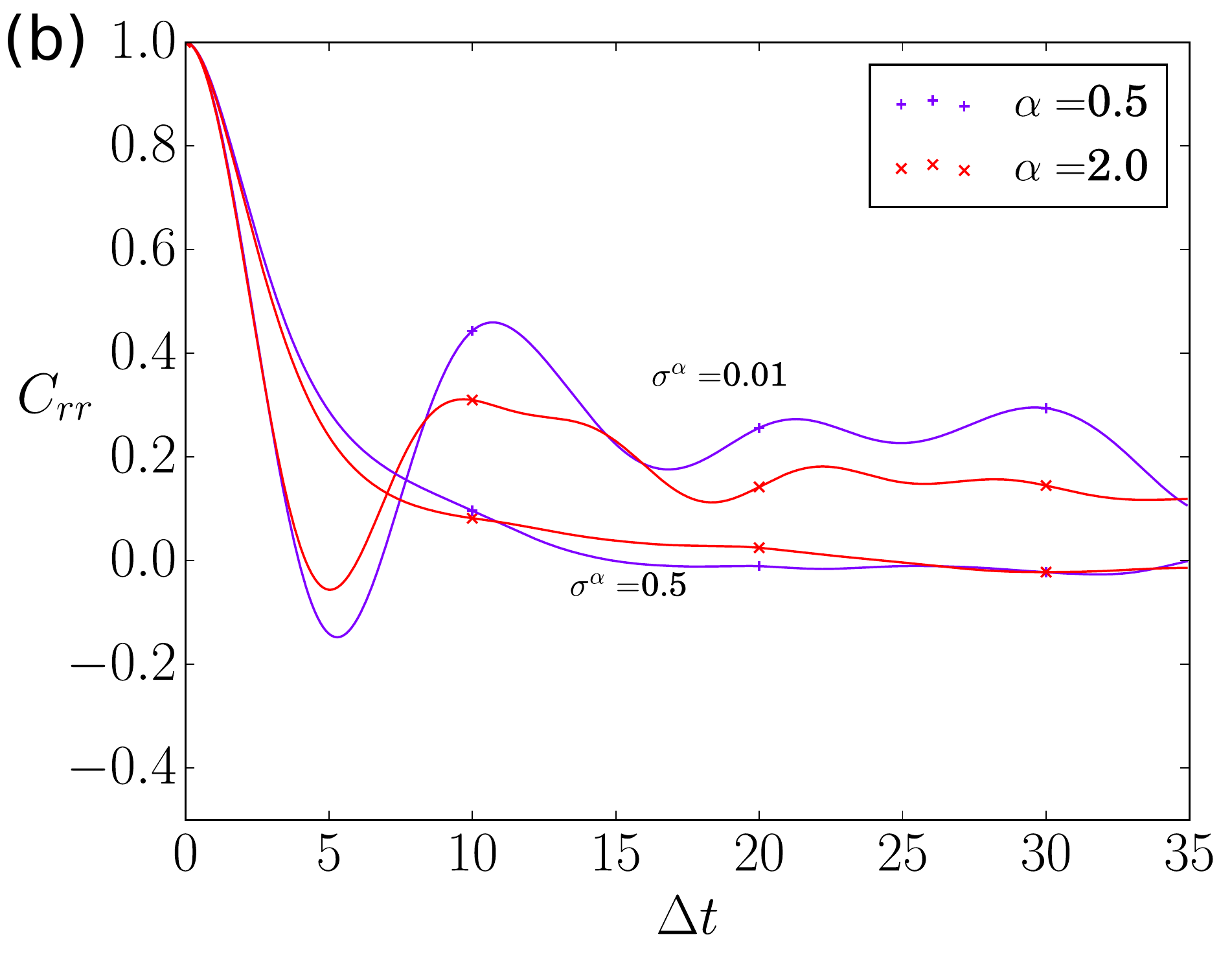}
    \caption{(a) Distance $r(t)$ from the home in dependence of time for two different noise types and two different noise strength. 
    (b) Corresponding autocorrelation function of distances $C_{r,r}(\Delta t)$. Values of $\alpha$ and $\sigma$ are given in the 
    graphs. $v_0=1$, $\kappa=1$.}
    \label{fig:return}
\end{figure}

Our searcher performs a quasiperiodic motion around the home. Figure \ref{fig:return} (a) shows the 
radial distance from the home for four sample trajectories. 
On the right the time dependent  autocorrelation function of distances corresponding to the trajectories 
on the left is plotted. In the simulations, we calculated
\begin{equation}
C_{rr}(\Delta t)=\lim_{T\to\infty}\frac{1}{T}\int_0^T (r(t)-<r>)(r(t+\Delta t)-<r>){\rm d}t\,.
\label{eq:crr}
\end{equation}
Further on we assumed stationarity and have set $<r>=2v_0/\kappa$ in agreement with \eqref{eq:p0}. 
We underline the good qualitative agreement of the presented results from simulations with the findings for the fruit fly 
\cite{Kim_Dickinson_2017,ElJundi_2017}. For comparing Figure \ref{fig:return} (b) with Figure 6.B of \cite{Kim_Dickinson_2017} we consider
$v_0=10mm/s$ as given by \cite{Kim_Dickinson_2017} as realistic value. With this value
the lag in $mm$ provided in Figure 6.B of \cite{Kim_Dickinson_2017} becomes a time scale if divided by the velocity.  

\begin{figure}[h]
    \includegraphics[width=0.6\linewidth]{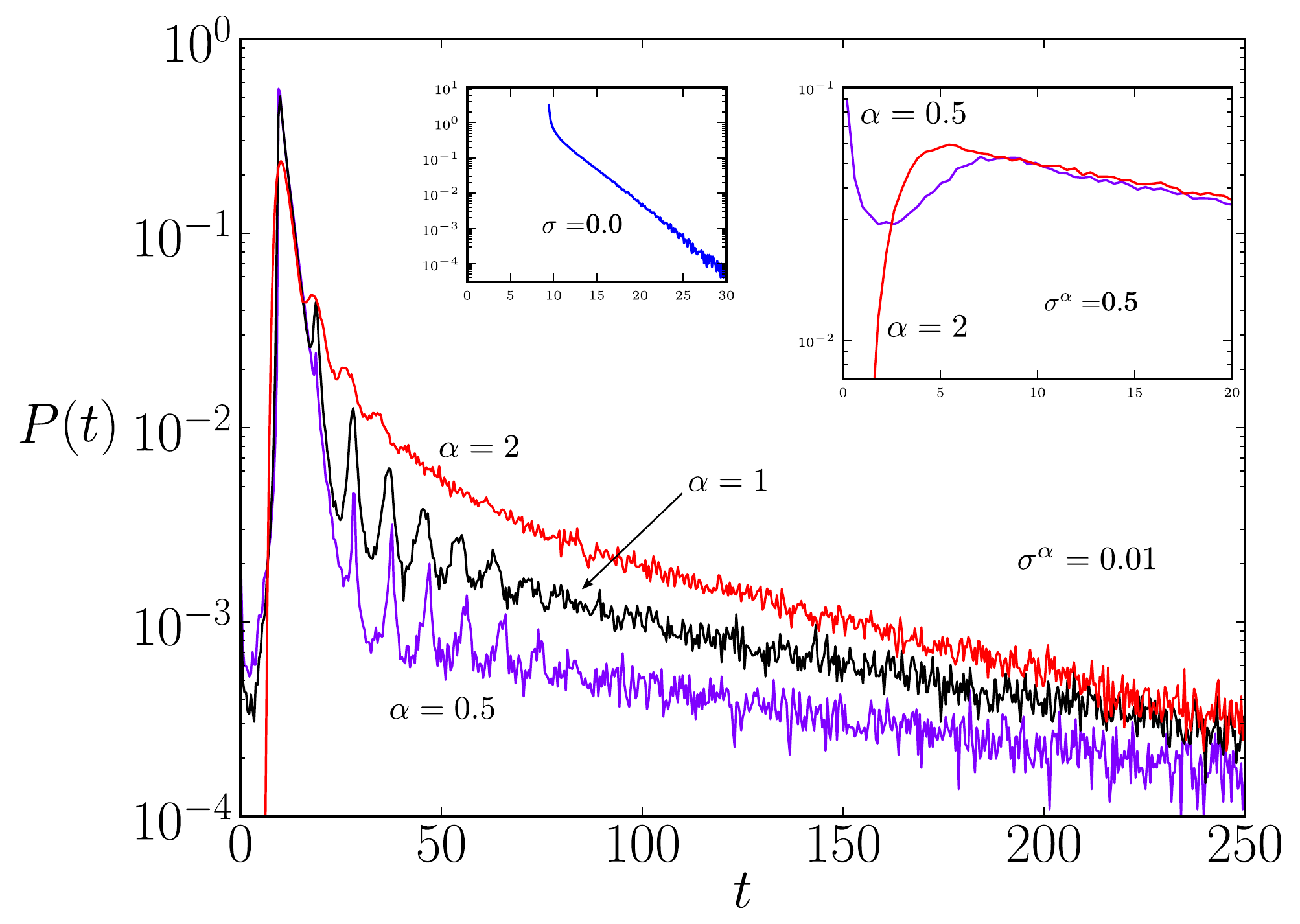}
    \caption{Return time to the home for three different noise types and three noise strength. The home is assumed to be extended. 
    Particles start at a distance $r(t=0)=0.1$ from the home and are considered to have returned when the distance is $r=0.1$ again. 
    For description of the insets, see in text.
    Other parameters: $v_0=1$, $\kappa=1$.}
    \label{fig:return2}
\end{figure}

In Figure \ref{fig:return2} the sojourn time distribution to a home of the size $r_s=0.1$ is shown. The particles start at the home with $r(t=0)=r_s$ and the initial 
angles $z_0$ are uniform distributed between $z_0 \in [0, \pi/2]$ and $z_0 \in [-\pi, -\pi/2]$. 
Those initial conditions correspond to being at the home and moving away from the home in the steady state. 
Note, that equidistribution of $z_0$ does not imply a uniform distribution of the $X$-values. 
The particle is considered to have returned to the home, if it touches the extended home for the first time, i.e. $r(t>t_0)=r_s$. 

The parameters are chosen to be consistent with the previous examples. The picture shows a low noise intensity case $\sigma^\alpha=0.01$, 
while the inset on the right
displays a rather large noise intensity $\sigma^\alpha=0.5$. The inset in the middle corresponds to the deterministic case. 
The speed is $v_0=1$ and the coupling to the home $\kappa=1$ is constant. 
The distribution of sojourn times for the deterministic case can be seen in the small inset. Note that
particles start returning to the home after $t_p\approx9.5$. The particles which return first correspond to the initial condition $r_0=r_s$ and $z_0=\pi/2$.
All other initial take a longer period of time. For $t>t_p$ the distribution strongly descends and takes an exponential form.

In the central picture sojourn time distributions are plotted for three noise types. All three noise types display a sharp peak at approximately $t_p\approx9.5$. 
This time $t_p$ is the time a deterministic particle with $r_0=r_s$ and $z_0=\pi/2$ needs for returning to the home $r(t)=r_s$.
The shape of the peak is influenced by the noise type, for the Gaussian noise $\alpha=2$ the peak is rather smooth compared to the Cauchy noise $\alpha=1$ and even more so 
for $\alpha=0.5$. For the lowest displayed noise the decay of the first peak approximately resembles the exponential decay of the deterministic particles. 

The behavior at the first peak can be understood by considering Figure \ref{fig:r_X_noise}. The left and the middle picture, corresponding to $\alpha=0.5$ and $\alpha=2.0$,
display the sample trajectories in the $(r,X)$ plane. Here the noise types act very distinct on the motion of the particles, while a Gaussian noise almost at 
every instant
in time causes a comparatively small change in the deterministic trajectory $X$, the other extreme, the $\alpha=0.5$ case, rarely cause a change, but if it causes a change,
the change can be large. This behavior causes here an almost deterministic shape of the first peak for $\alpha=0.5$ and the smearing out of the peak for Gaussian white noise.
Before the peak the different noise distributions also cause distinct behavior. For Gaussian noise basically no particle returns for times roughly $t<5$, while in the other 
cases some particles immediately return. As L\'evy noise with smaller $\alpha$ increases the probability of large sudden changes, this different behavior can be understood.
For non-Gaussian noise the direction of motion of the particles can suddenly jump by $\pi$ and therefore allowing an immediate return of the particle. 

After the first peak several other peaks follow in all cases, but with different intensity. In fact, already in the first peak a second peak is visible. Those peaks correspond
to multipliers of $nt_p$ $n\in N$. Best visible are those peaks for $\alpha=0.5$ up to roughly the correlation time $\tau$.  

A high noise intensity causes almost the same shape of the time distribution for all noise types. Differences only remain at small times $t\ll t_p$, as can be seen 
in the inset on the right. Both curves seem to decay in the same way and therefore only a small time interval is shown.

\subsection{Mean first hitting time}
\label{subsec:meanfirsthit}
\begin{figure}[h]
    \includegraphics[width=0.5\linewidth]{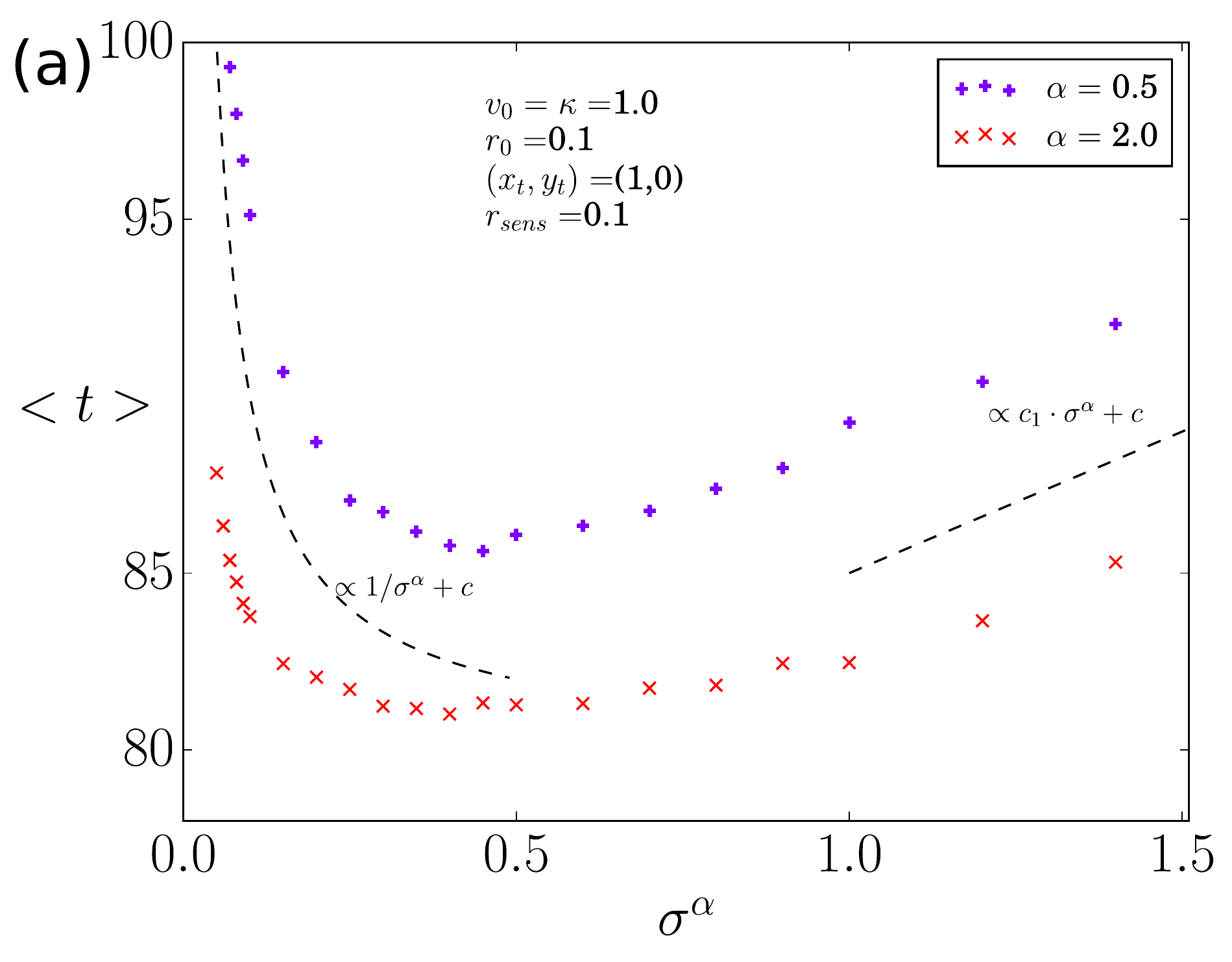}
    \includegraphics[width=0.48\linewidth]{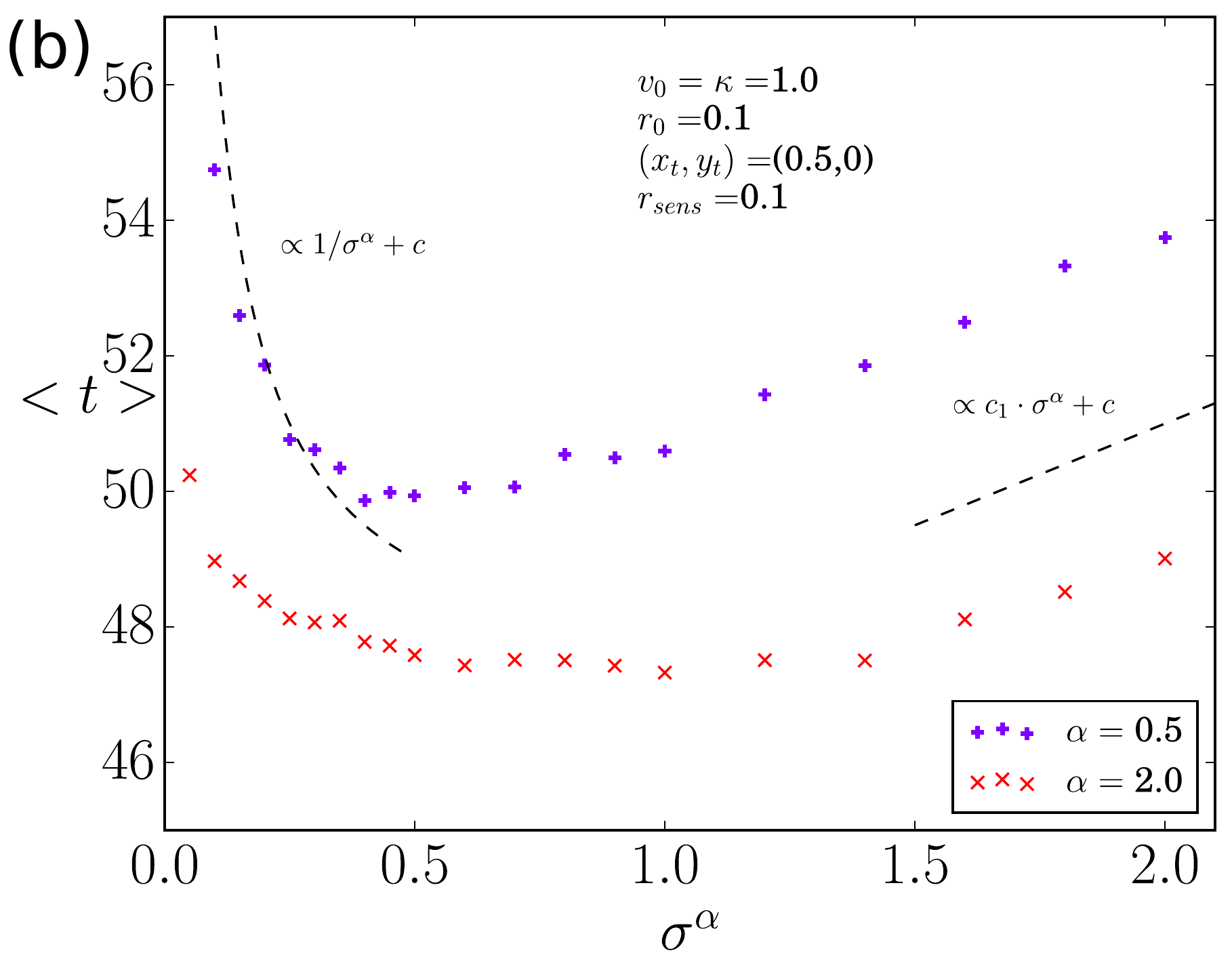}
    \caption{Mean time $<t>$ till a new spot $(x_t,y_t)$ at a given position in the $(x,y)$ 
    plane is found in dependence of the noise intensity. (a) $(x_t=1,y_t=0)$ (b) $(x_t=0.5,y_t=0)$.} 
    \label{fig:meet}
\end{figure}
For insects an oscillatory motion around a given home is often considered to be foraging, the question arises how fast can a particle 
in our model discover a food source. We assume our searcher has a sensing radius $r_{\text{sens}}$.
This sensing radius is small against the length scale $r_c$ of our system, we chose $r_{\text{sens}}=0.1\ll r_c=1$. We expect
the sensing radius to be small against $r_c$ as otherwise not much new space can be discovered. We performed 
simulations of particles starting close to the home $r(t=0)=r_0=r_{\text{sens}}$, with uniformly distributed heading directions and position angles
and determined the mean first hitting time $<t>$ 
till the food source is discovered. The localized target or the new food source was placed at $(x_t,y_t)=(1,0)$ and at $(x_t,y_t)=(0.5,0)$. 
Localized means, the target is point like. We chose the first
spot at a distance of $r_c$, as every deterministic bounded trajectory crosses this distance 
and the spatial distribution  \eqref{eq:p0} possesses maximal probability, there. As second
spot, we chose a closer one, to investigate how the behavior of the mean first hitting time changes. The searcher usually returns several times
to the home till the food source is discovered.
Figure \ref{fig:meet} shows the mean first hitting time $<t>$ for a given new food spot $(x_t,y_t)$ in dependence of the inverse relaxation time $1/\tau$. 
Since we have fixed $v_0=1$ the latter is according to Equation \eqref{eq:tau} proportional to the scaled noise intensity $\sigma^\alpha$.

In both pictures an optimal noise strength can be seen, at which the mean first hitting time $<t>$ is minimal. 
This optimal time does depend on the relaxation time $\tau$ but also on the distance between 
the food source and the home. The optimal noise strength depends on the distance where the target is situated and decays with growing distance.

The non-monotonous dependence of the hitting times can be explained as resulting from two counteracting effects originated by the noise. 

Considering a deterministic motion $\sigma=0$ not every searcher will hit the target as most of the radial unbounded trajectories will miss it.
For bounded trajectories the mean time will be proportional to the period length. We remind here, that the sensing radius is small, so even after
the accumulated apsidal precession is larger than $2\pi$ the particle is likely not to have found the food source. 
In the deterministic case the mean first hitting time diverges.
With small noise present the searcher can switch deterministic trajectories. Now particles following 
a deterministic trajectory that never hits the target, or that takes extremely long to return, might arrive earlier through switching. 
This redistribution is proportional to the time scale of 
the integral of motion $X$ given by \eqref{eq:avXt}. Correspondingly the mean first hitting time decays proportional to $\tau$ from \eqref{eq:tau}. 
The mean first hitting time is always larger than the relaxation time $\tau$. 


The second counteracting process starts to act at a higher noise. 
Now, the deterministic motion can be practically neglected and slow diffusive motion of the searcher dominates. The time scale
for diffusive motion is given by \eqref{eq:fpe_over} and \eqref{eq:deff_search}. 
Therefore, the time after which a certain distance $\Delta r$ has 
reached in average by this diffusion can be estimated as  
\begin{equation}
t_{\rm diff} = \frac{<\Delta r^2 >}{4 D_{\rm eff}}=\frac{\sigma^\alpha}{2 v_0^{2+\alpha}} <\Delta r^2>\,.
\label{eq:diff}
\end{equation}
Thus, for large noise intensity the hitting time starts growing $\propto\sigma^\alpha$. Since the latter scales as $\propto 1/\tau$ we consolidated the second noise induced effect governing the increasing hitting times.

In our numerical simulations Gaussian white noise always performed slightly better as all other noise types $(\alpha<2)$.
When changing the noise parameter $\alpha$ to lower values we observed only a small increase of the mean time $<t>$.
So, different turning statistics seem not to significantly improve the local search in our model, if $<t> \ge \tau$. 

This result is contrary to a result for freely moving Daphnia
\cite{Garcia_daphnia} during global search, where an optimal turning angle distribution for search was found. 
We will show in a follow up work, that however for a searcher with an uncertainty 
of the exact position angle the turning angle distribution significantly can influence the success of returning home.

\begin{figure}[h]
    \includegraphics[width=0.49\linewidth]{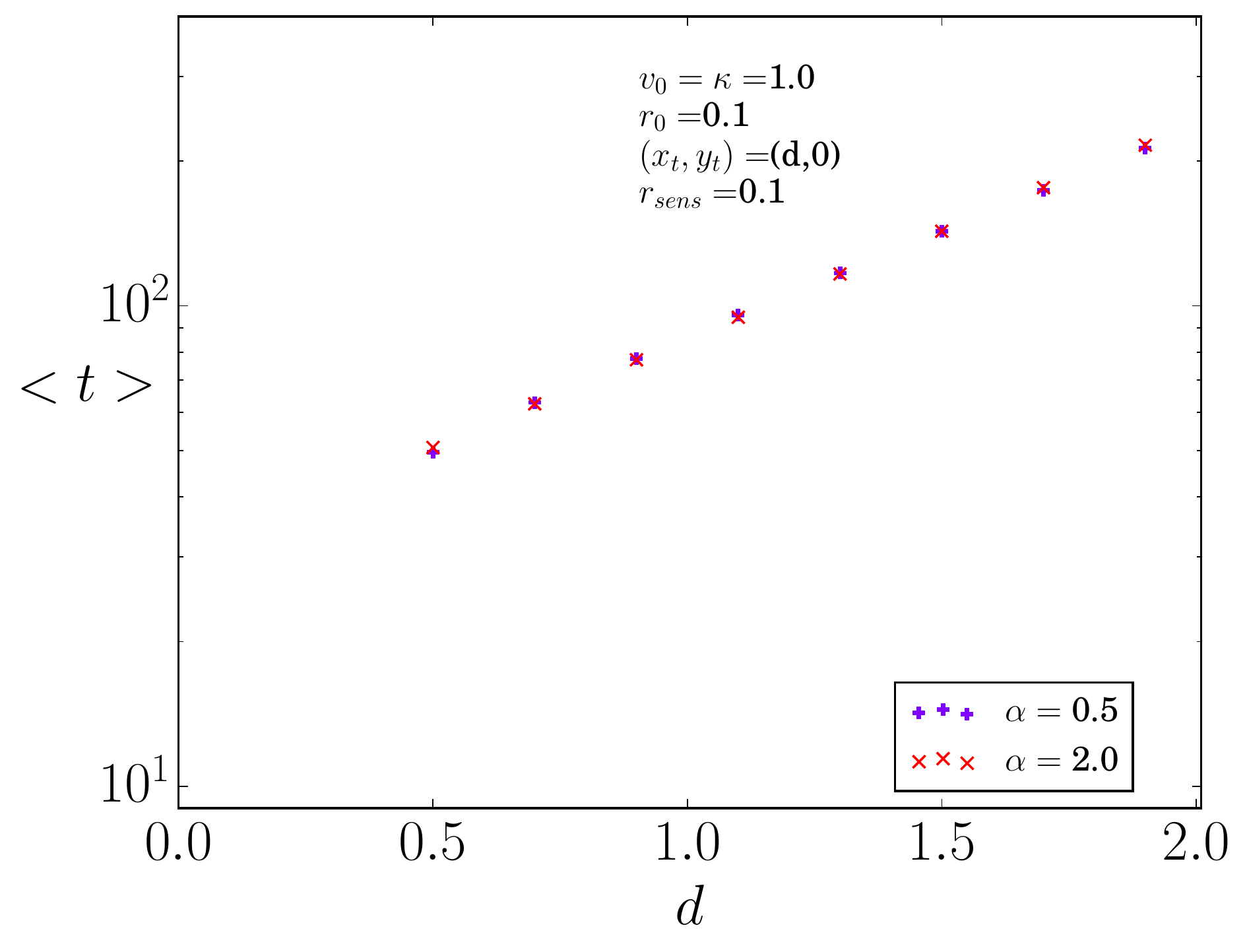}
    \caption{Mean time $<t>$ for a fixed noise intensity $\sigma^\alpha=0.4$ 
    in dependence of the spot distance $d$, with $(x_t=d, y_t=0)$. } 
    \label{fig:meet2}
\end{figure}

In Figure \ref{fig:meet2} the mean first hitting time for fixed noise strength $\sigma^\alpha=0.4$ in 
dependence of the spot distance $d$ is shown. As can be seen, the times grow exponentially fast with the distance. We point out 
that for each of the considered distances we observed a non-monotonic dependence of the mean hitting times  with respect to the 
noise intensity $\sigma^{\alpha}$.  

We mentioned in section \ref{sec:spat_dist} that we consider $v_0=10mm/s$ and $\kappa=0.5/s$ and therefore $r_c=20mm$ as realistic values 
for a fly. We mention here, that for such values and a food source at a spot distance $d=20mm$ the optimal time becomes $<t>=160s$ and with $d=40mm$ the optimal 
time becomes $<t>=800s$.


\section{Coupling dependent on distance}
\label{sec:extension}
In this Section, we generalize the model through replacing the
coupling parameter $\kappa$ to the home by a distance dependent coupling
$\kappa(r)$, with $r=\sqrt{x^2+y^2}$, when considering Cartesian
coordinates. By doing so, we cover a broader class of spatial
distributions and deterministic trajectories and the model might describe also more complex motions with preferred paths. It is straight forward
to adjust the so far determined properties: the deterministic
trajectories \eqref{eq:sinzvr}, the spatial distribution
\eqref{eq:p0}, the relaxation time \eqref{eq:tau}.

We introduce the time evolution of the direction of the velocity
$\vec{v}$
\begin{equation}
\dot{\theta} =  \kappa(r) \sin(\theta-\beta) + \frac{\sigma}{v_0} \xi(t),
\label{eq:dotthetaU}
\end{equation}
with coupling $\kappa(r)$. We require that the function $\kappa(r)$ has no singularity for all distances including $r=0$.
This way we ensure that at the origin $(x,y)=(0,0)$ the system in Cartesian coordinates is well defined. 
In addition to ensure the existence of a steady state, the spatial density of searchers shall be normalizable. This will set 
another condition on $\kappa(r)$.

\begin{figure}[h]
    \includegraphics[width=0.4\linewidth]{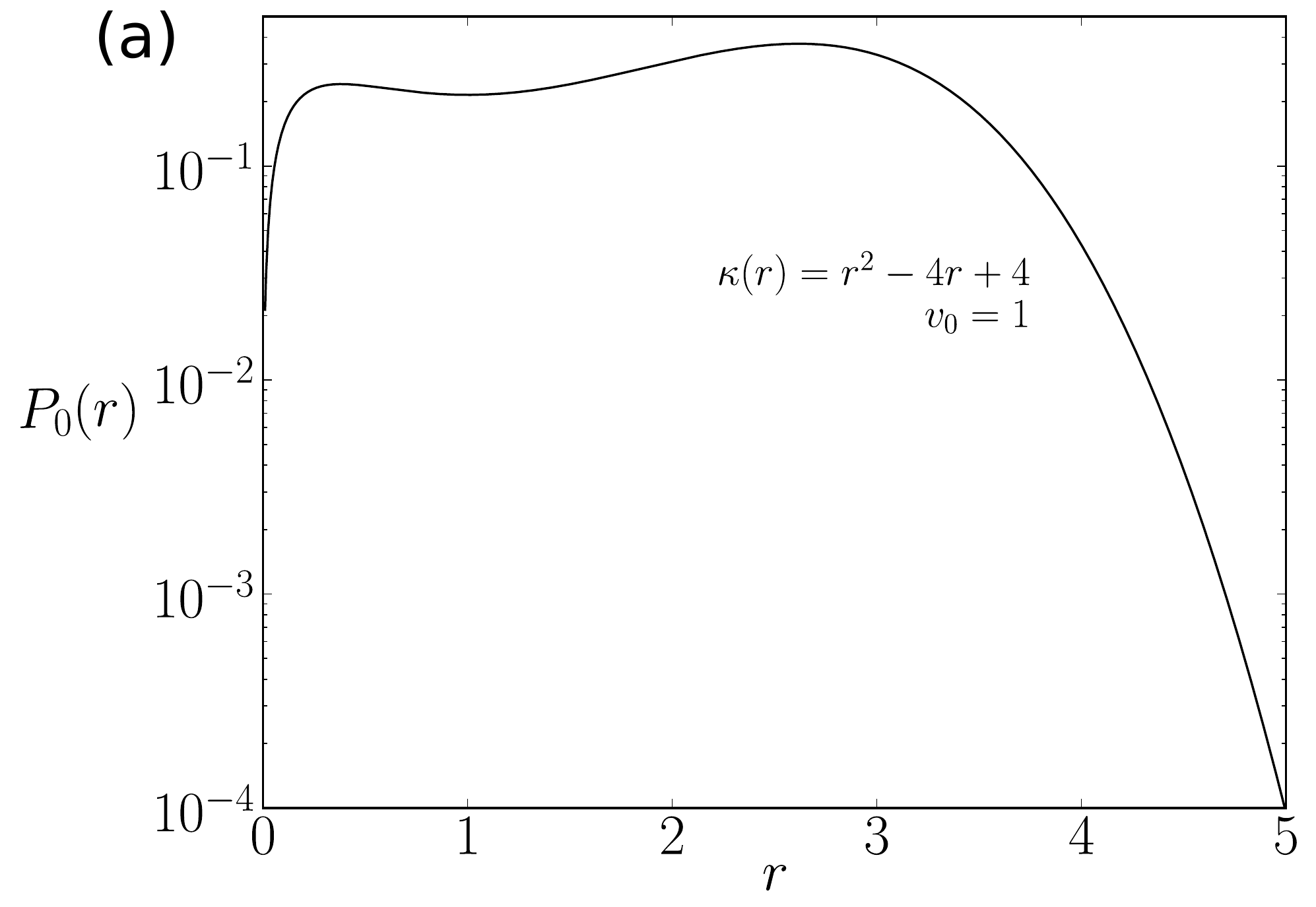}
    \includegraphics[width=0.44\linewidth]{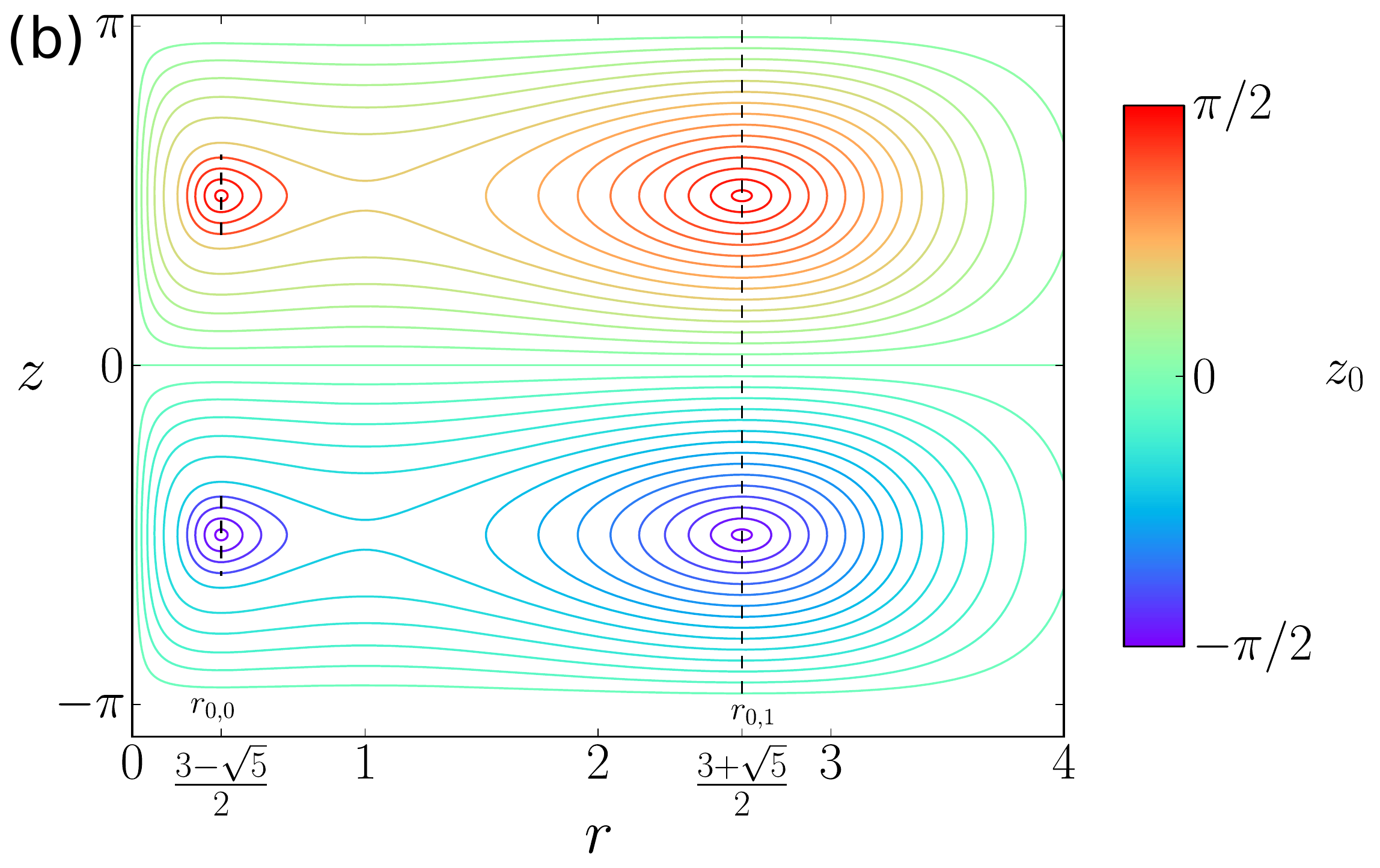}
    \includegraphics[width=0.34\linewidth]{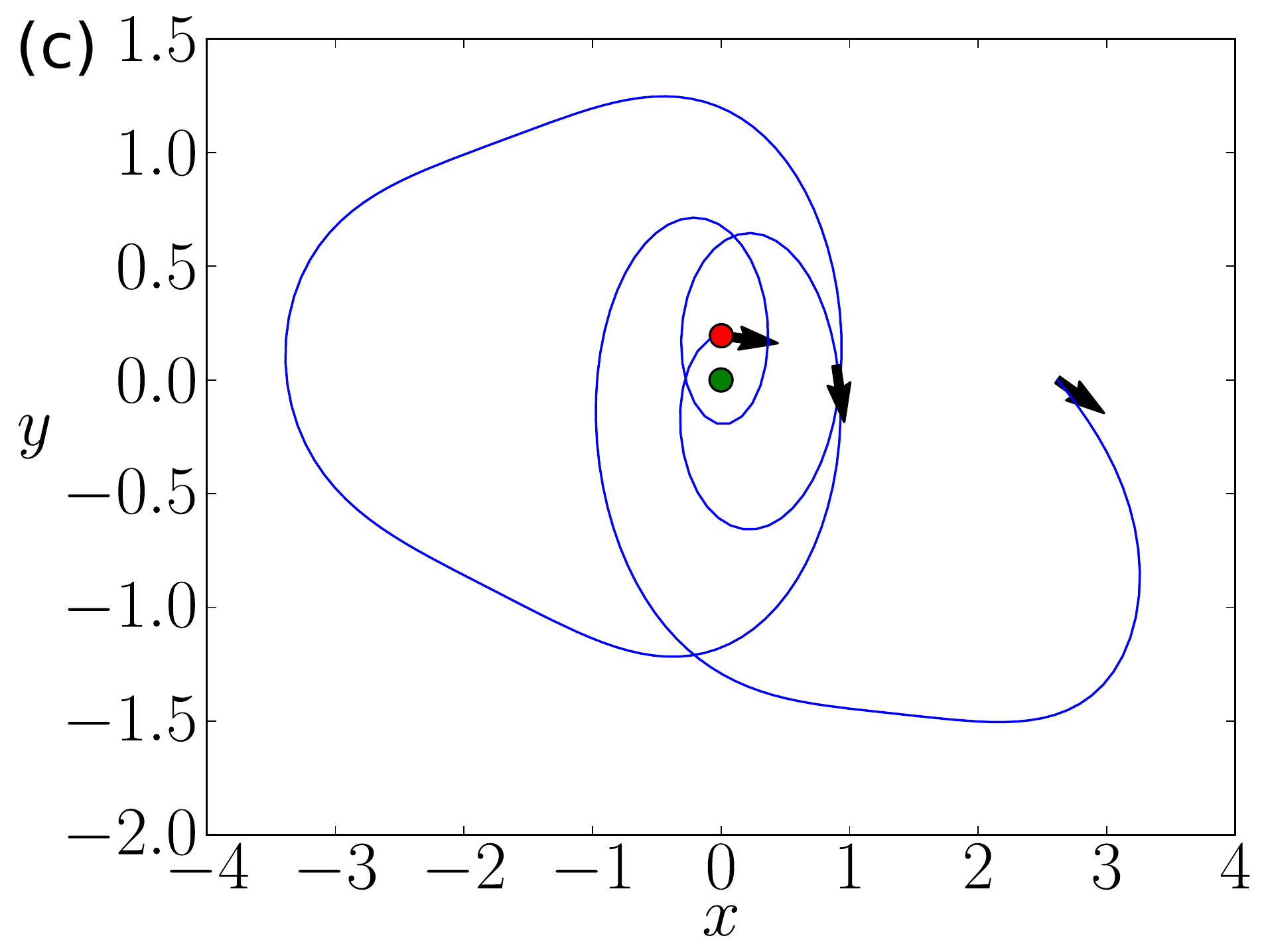}
    \includegraphics[width=0.45\linewidth]{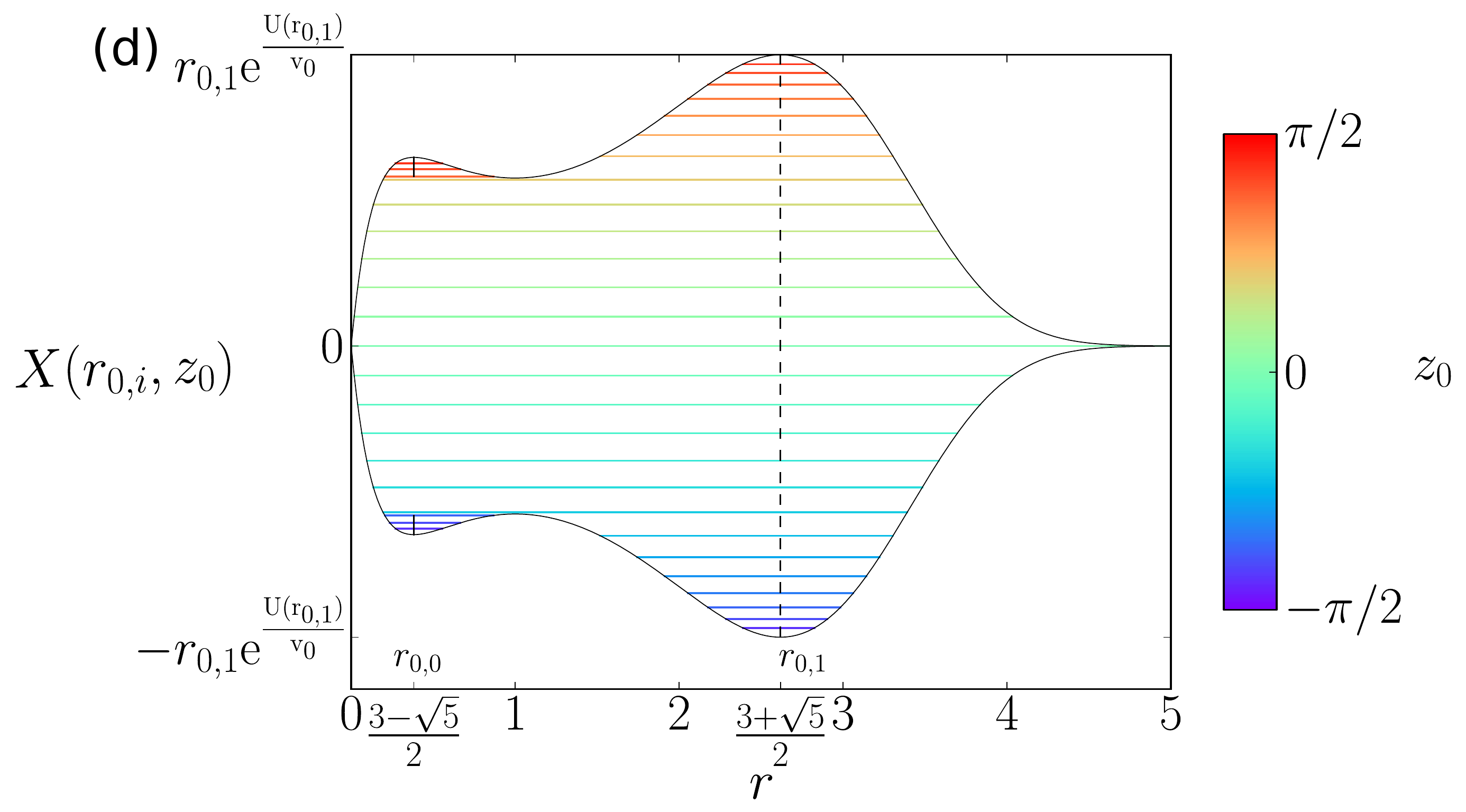}
    \caption{Example for distance dependent coupling strength $\kappa(r)$. (a) Steady state spatial distribution $P_0$. 
    (b) Corresponding deterministic trajectories in the $(r,z)$ plane, with initial condition
    $r_0$, as indicated by broken lines and initial angle $z_0$ according to the colorbar. 
    (c) Sample trajectory in the $(x,y)$ plane without noise corresponding to the blue separatrix in the $(r,z)$ plane. 
    (d)
    Trajectories in the $(r,X)$ plane with initial condition $r_0$, as indicated by broken lines and initial angle $z_0$ according to the colorbar.}
    \label{fig:capU}
\end{figure}

The time evolution of the angle $z$ between the direction of motion $\theta$ and the position of the home $\beta$ is now given by:
\begin{equation}
\dot{z} =  -\left(\frac{v_0}{r}-\kappa(r)\right) \sin(z). 
\label{eq:dotzU}
\end{equation}
It follows for the deterministic trajectories, that  
\begin{equation}
\sin(z(r))\exp\left(-\frac{U(r)}{v_0}\right)r=X=\sin(z_0)\exp\left(-\frac{U(r_0)}{v_0}\right)r_0                  
\label{eq:sinzvrU}
\end{equation}
holds wherein we have defined
\begin{equation}
U(r)=\int^r {\rm d} r^\prime \kappa(r^\prime).
\label{eq:kappaofr}
\end{equation}
Again a particle with an initial angle $z_0=0$ or $z_0=\pi$ will move on a straight line. There also exists a minimal and a maximal distance from the home, with $\sin(z)=\pm1$.
Previously, in Equation \eqref{eq:sinzvr} we could fix $r_0$ such that the integral of motion $X(z_0)$ only depended on the initial angle $z_0$. 
The value of $X(z_0)$ represented a trajectory in the $(r,z)$ plane, or in the $(r,X)$ plane in an unique way. This might now be no longer true. Now the 
variable $X(r_0,z_0)$ might be dependent on the initial position and the initial angle, if one wants to uniquely identify a trajectory in the $(r,z)$ plane and have $z_0$ as parameter. One could, 
however, fix the initial angle
$z_0=\pm\pi/2$ and then the variable $X(r_0)$ would still be dependent on only one parameter and be unique for the trajectories in the $(r,z)$ plane, 
but we chose the initial angle $z_0$ to be a parameter. 
Figure \ref{fig:capU} displays such a case. The upper left plot 
shows the steady state pdf. It is obtained from \eqref{eq:fpe_full} by replacing $\kappa$ with $\kappa(r)$ and following
the same steps as in Section \ref{sec:spat_dist}:
\begin{equation}
P_0(r,z)=cr\exp\left(-\frac{U(r)}{v_0} \right),
\label{eq:p0U}
\end{equation}
which again is independent of the noise. It reads in Cartesian coordinates
\begin{equation}
P_0(x,y)=c\exp\left(-\frac{U(\sqrt{x^2+y^2})}{v_0} \right),
\label{eq:pxy0U}
\end{equation}
with $c$ being the normalization constant. This sets the second condition on $\kappa(r)$ as we require the normalization to be possible. 

The noise still acts perpendicular on the deterministic trajectory $X$ causing a switching and the ensemble average \,$<X>$\, still follows the exponential decay
from Equation \eqref{eq:avXt}, with the same relaxation time as given before \eqref{eq:tau}. 
The overdamped dynamics for the radial transition pdf $P(r,t|r_0,t_0)$
\begin{eqnarray}
&&\frac{\partial}{\partial t}P=D_{{\rm eff }} \frac{\partial}{\partial r}\left(\frac{\partial}{\partial r}P-\left(\frac{1}{r}-\frac{\kappa(r)}{v_0} \right)P\right)\,,
\label{eq:fpe_overU}
\end{eqnarray}
has \eqref{eq:p0U} as asymptotic steady pdf. As above, the radial relaxation again slows down with increased noise strength. 
Likewise in the situation with constant $\kappa$, only $\tau$ or $D_{{\rm eff}}$ expresses the influence of the various $\alpha$-values of the noise.

Figure \ref{fig:capU} gives an example for a space dependent coupling $\kappa(r)$. We chose $\kappa(r)=r^2-4r+4$ such that the steady state pdf exhibits two maxima. 
This can be seen in panel (a). 
In panel (b), the deterministic trajectories are shown in the $(r,z)$ plane with two spatial initial condition $r_{0,i}$ ($i=0,1$) 
as indicated by the two broken lines and initial angle $z_0$ according to the colorbar are shown. 
Due to the quadratic term in the coupling trajectories are significantly shorter than in the initial model since the coupling to the home 
pointing direction increases with the distance from the home. 

Extremal points of radial probability follow from ${\rm d}P_0/{\rm d}r=0$, respectively, for the considered coupling from  
\begin{equation}
1-\frac{r}{v_0} \kappa(r)=0. 
\label{eq:ext}
\end{equation}
Solutions are easily found. Minimal probability is found around $r_1=1$ and maximal one at the two distances $r_{2,3}=(3\pm \sqrt{5})/2$ 
as shown in Figure \eqref{fig:capU} (a). In the $r,z$-plane (see (b) of the same Figure, these distances 
are connected with the fixed points of the deterministic flow of trajectories located at angles $z=\pm\pi/2)$.  
The fixed points at $r_1$ are of saddle type whereas these, at the two other distances, are centers. 
They always correspond to a circular solution for the deterministic trajectories in the $(x,y)$ plane since at those points 
the angular and the radial velocity $\dot{z}=0$, $\dot{r}=0$ vanish.

Small changes in the initial angle $z_0$ around the circular solutions in the $(x,y)$ plane at the maxima $r_{2,3}$ correspond to 
trajectories in the $(x,y)$ plane similar to Figure \ref{fig:x_y_plane} (a).
For initial angles $z_0$ close to $0$ or $\pm\pi$ trajectories in the $(x,y)$ plane will be comparable to  
Figure \ref{fig:x_y_plane} (b), 
but shorter with faster turnings. Between the two mentioned solutions a separatrix lays. 
The picture (c) is a trajectory in the $(x,y)$ plane close to the separatrix. 
This solution exists if the initial angle $z_0=\pm\pi/2$ at the minima is slightly changed. 
The resulting trajectory in the $(x,y)$ plane rotates two times around the home before reaching the maximal distance twice. The noise facilitates again switchings between trajectories.

If the steady state pdf of the position can be experimentally measured it can be fitted to our solution \eqref{eq:pxy0U}, thus determining the coupling strength $\kappa(r)$ to the home. 
Having found a suitable dependence, the relaxation time $\tau$ can be determined by the ensemble or time average of the variable $X$ through measuring $r(t)$ and $z(t)$. Fitting experimental data thus allows to determine the relaxation time $\tau$, and therefore the noise strength $\sigma^\alpha$ of the model.

\section{Conclusions}
\label{sec:concl}
We laid the foundation for a minimal stochastic model for a local searcher 
which was motivated by experimental observations of the stochastic oscillatory 
motion of insects around a given home. 
The main ingredients of this minimal model are the constant speed of the searcher and stochastic angular variation, 
that only requires the knowledge 
of the position angle and the heading direction, which allows the 
particle to explore the vicinity of the given home. The specific 
interaction with this home results in an exploration of the neighborhood 
around the home and an attraction towards it in dependence on the mutual orientations of the position and heading vectors. 

The model was formulated with four parameters. $\kappa$ defines the strength of interaction with the home, $v_0$ 
is the speed of the searcher and $\sigma$ stands for the intensity of the noise. 
Since the observed turning angle behavior in experiments such as fruit flies can be of Non-Gaussian statistics 
we introduced $\alpha$ stable noise as source of randomness. The corresponding parameter $0<\alpha \le 2$  
as fourth parameter allows to vary the support of the noise source between special types of noise as Gaussian, Lorentzian, etc.

The introduced model showed qualitative agreement with the behavior of insects. The advantage of our model is the analytical and simple numerical tractability. 
In consequence we were able to discuss typical behavior of the  trajectories and of characteristic times. 
For example, we found the characteristic return times in the noise free case and obtained an apsidal precession 
of the oscillatory trajectories reminiscent of celestial motion. The analysis of the models allowed us to discuss in detail the 
deterministic properties and the effects originating through the addition of different symmetric white noise sources. 

The inclusion of noise has a stabilizing effect on the system since unstable trajectories disappear. Generally,  
trajectories  start to randomize. This is manifested by the noise dependent relaxation time $\tau$ that is proportional to $1/\sigma^\alpha$. 
For larger times the stochastic dynamics has forgotten its initial directions and trajectories have spread over all possible orbits. 
This investigation has concentrated on the relaxation of the deterministic integral of motion $X$. Its first moment conditioned to 
initial values has decayed at times larger $\tau$. 

At high noise the particles start to perform diffusive motion. As every active stochastic particle, 
the corresponding effective diffusion coefficient depends inversely proportional on $\sigma$. 
We derived for the nonlinear model the overdamped Smoluchowski equation outgoing from the FPE for all $\alpha$-values. 
It describes the stochastic dynamics on times scales much larger than the noise dependent time and $\tau$. Except from $\tau$ the Smoluchowski 
equation is independent of the noise characteristics.  

We obtained analytically the steady state spatial distribution $P_0$ which appeared to be identical for all different kinds of $\alpha$-stable noise, see Equation \eqref{eq:p0}.  
In particular, it does depend neither on $\sigma$ nor on $\alpha$. 
Distances are exponentially distributed and the width is determined by the ratio $v_0/\kappa$ which is just a characteristic length for the quasiperiodic excursions which the stochastic trajectories perform. 

We found in our model an optimal noise strength for finding a new spot in minimal average time $<t>$. 
This optimal average time is distance dependent. The searcher finds on average the new spot always faster with noise in the angular dynamics. 
This is the result of the relaxation towards a probabilistic population of all possible trajectories which determines the greater success of the stochastic searcher. For lower noise 
this process is governed by the noisy periodic motion and after the relaxation time the stationary pdf is established.  However, for larger noise the relaxation is proceeded by diffusive search. If an approximate distance of a spot to discover is given, 
we expect a good choice that this length equals the length scale $r_c$ of our system. We found in our model only an insignificant dependence on the turning angle distribution, expressed through the noise type. This result is in contrast to results for global search \cite{Garcia_daphnia}. We will show however in a future publication that the probability
to return to the home does strongly depend on the noise type if an uncertainty of the position angle exists.

Another advantage of our model is that we could generalize the model to distance dependent coupling $\kappa(r)$ and thus allowing to express a large class of radial symmetric spatial steady state distributions, see Equation \eqref{eq:p0U} and corresponding spatial trajectories. All models of this class exhibit the time scale $\tau$ so that we expect the existence of an optimal noise strength for the general model.
The long time behavior of systems with distance-dependent coupling follows again a Smoluchowski equation uniformly for all $\alpha$'s.

We underline that our findings are applicable to a broad class of stochastic searching units as insects 
as well as  autonomous vehicles. Here, we considered individual searchers. The investigation concentrated on the interaction with the given home and on temporal scales to find new food sources. We will show in a follow up work, 
that an uncertainty 
of the exact position angle can significantly influence the turning angle distribution and, therefore, the success of returning home. 
Further research on interacting searchers and their cooperative behavior is in progress.     

\section{Acknowledgments}
This work was supported by the Deutsche Forschungsgemeinschaft via
grant IRTG 1740 and by the Sao Paulo Research Foundation (FAPESP) via
grants 2015/50122-0 and 2017/04552-9.  LSG thanks Alexander
Neiman and Ohio University in Athens OH for hospitality and support.
The authors thank Fabian Baumann for fruitful discussions.

\appendix

\section{Derivation of stochastic $X$-dynamics}
\label{app:X-dyn}
Here we derive the linear differential equation for the averaged value of $X$ as defined by 
\begin{equation}
\int_0^\infty \text{d} r\int_{-\pi}^{\pi} \text{d} z\, X(r,z)\,P(r,z,t|r_0,z_0,t_0)\,.
\label{b1}
\end{equation}
(comp. Equation \eqref{eq:aveX}). This characterizes the relaxation of the  stochastic $X(r,z,t)$ dynamics. 
It is dissipative in contrast to the deterministic behavior. 
To obtain the dynamics for the average we multiply the FPE \eqref{eq:fpe_full} for the transition pdf $P=P(r,z,t|r_0,z_0,t_0)$ by $X$ 
from\eqref{eq:sinzvr} and integrated over $r$ and $z$:
\begin{eqnarray}
&&\int_0^\infty \text{d} r\int_{-\pi}^{\pi} \text{d} z\, X(r,z)\,\frac{\partial}{\partial t}P=\\
&&=\int_0^\infty \text{d} r\int_{-\pi}^{\pi} \text{d} z\, X(r,z)\left[-v_0\frac{\partial}{\partial r}\cos(z)+\frac{\partial}{\partial z}\left(\frac{v_0}{r}-\kappa \right)\sin(z)+\left(\frac{\sigma}{v_0}\right)^\alpha \frac{\partial^\alpha}{\partial |z|^\alpha}\right]P\,.\nonumber
\end{eqnarray}
The first term at the r.h.s after partial  integration over $r$ together with the second term partially integrated over $z$ 
results in an expression which vanishes if $X(r,z)$ is inserted.
\begin{eqnarray}
\int_0^\infty \text{d} r\int_{-\pi}^{\pi} \text{d} z\,\left[v_0\cos(z)P\frac{\partial}{\partial r}X-\left(\frac{v_0}{r}-\kappa \right)\sin(z)P\frac{\partial}{\partial z}X \right] =0\,.
\end{eqnarray}
Thus we are left with :
\begin{eqnarray}
\frac{{\rm d}}{{\rm d} t}<X> =\left(\frac{\sigma}{v_0}\right)^\alpha\int_0^\infty \text{d} r\int_{-\pi}^{\pi} \text{d} z\, X\frac{\partial^\alpha}{\partial |z|^\alpha}P(r,z,t|r_0,z_0,t_0)
\label{b3}
\end{eqnarray}
We express the pdf through its Fourier transform
\begin{equation}
P(r,z,t|r_0,z_0,t_0)=(2\pi)^{-1}\int_{-\infty}^\infty {\rm{ d}}k\exp(-ikz)P(r,k,t|r_0,z_0,t_0)\,. \nonumber 
\label{eq:fourier}
\end{equation}
Its introduction in \eqref{b3} and taken the angle $z$ to be unwrapped yields:
\begin{equation}
\frac{{\rm d}}{{\rm d}t} <X> =-\frac{1}{2\pi}\left(\frac{\sigma}{v_0}\right)^\alpha\int_0^\infty \text{d} r\int_{-\infty}^{\infty} \text{d} z\int_{-\infty}^\infty {\rm{ d}}k\exp(-ikz)\, X(r,z)|k|^\alpha P(r,k,t|r_0,z_0,t_0)\nonumber
\end{equation}
Including the definition of $X$ and performing the $z$ integration leads to:
\begin{eqnarray}
&&\frac{{\rm d}}{{\rm d}t} <X>=\nonumber\\
&&-\left(\frac{\sigma}{v_0}\right)^\alpha \frac{1}{2\pi}\int_0^\infty \text{d} r \int_{-\infty}^\infty {\rm{ d}}k\frac{1}{2{\rm{i}}}(\delta(k-1)-\delta(k+1))\, r
\exp\left(-\frac{r}{r_c}\right)|k|^\alpha P(r,k,t|r_0,z_0,t_0)\,,
\label{eq:Xrechts}
\end{eqnarray}
with the $\delta$-functions originated by the $\sin$-function in $X$. Eventually,  we have a look at the definition of the average \eqref{b1}. 
Therein we also perform  the Fourier transform and and take an unwrapped angle $z$ which gives:
\begin{equation}
<X>=\frac{1}{2\pi}\int_0^\infty \text{d} r\int_{-\infty}^\infty {\rm{ d}}k\frac{1}{2{\rm{i}}}(\delta(k-1)-\delta(k+1))\, r
\exp\left(-\frac{r}{r_c}\right)\,P(r,k,t|r_0,z_0,t_0)\,,
\label{eq:Xlinks}
\end{equation}
Comparing Equations \eqref{eq:Xrechts} and \eqref{eq:Xlinks} shows that after performing the $k$ integration both equations are 
identical up to the factor in front of \eqref{eq:Xrechts}, therefore it follows:
\begin{equation}
\frac{{\rm d}}{{\rm d} t}<X>=-\frac{1}{\tau} <X>
\end{equation}
This equation is solved by \eqref{eq:avXt} with $X_0$ being the initial value of $X$. 

One finds the same relaxation time also by using the time dependent eigenfunction of the FPE \eqref{eq:fpe_full}. It is solved by  
\begin{equation}
e_1(r,z,t) \propto \sin(z)r^2\exp\left(-2\frac{\kappa}{v_0}r-\frac{t}{\tau} \right)\,, 
\label{eq:efewm1}
\end{equation}
with eigenvalue $\lambda_1=1/\tau$. It describes the relaxation of the integral of motion $X$ in case 
that $\alpha$-stable noise is present in the angular dynamics.

\section{Derivation of the overdamped Smoluchowski-equation}
\label{app:r}
Following \cite{sevilla_telegraph, Sevilla_smol}, who discussed an overdamped description 
of freely diffusing active particles with Gaussian white noise, we define the 
Fourier components
\begin{equation}
P_n(r,t)=\int_{-\pi}^{\pi}{\rm{d}}z \exp(inz)P(r,z,t)\,, ~~~~~~n=0,\pm 1,\pm 2, \ldots
\end{equation}
For simplicity we omit the initial states in the transition pdf $P$ and in the components. Note, that the zeroth component with $n=0$ equals the marginal spatial pdf  $P(r,t)$, for which we want to derive an approximative equation as well as to get the necessary conditions for its validity.

We multiply the FPE \eqref{eq:fpe_full} from the left with $\exp(inz)$ and integrate over $z$. We obtain a set of coupled partial differential equations for the Fourier amplitudes:
\begin{equation}
\frac{\partial}{\partial t}P_n=-\frac{v_0}{2}\frac{\partial}{\partial r}\left(P_{n+1}+P_{n-1}\right)-
\frac{n}{2}\left(\frac{v_0}{r}-\kappa \right)\left(P_{n+1}-P_{n-1}\right)+
\left(\frac{\sigma}{v_0}\right)^\alpha  |n|^\alpha P_n\,.
\label{eq:fpe_pn}
\end{equation}
Afterwards, we eliminate the last term on the r.h.s. by substituting $P_n=\exp(-|n|^\alpha t/\tau)P^\prime_n$, with $\tau$ from \eqref{eq:tau}: 
\begin{eqnarray}
&&\frac{\partial}{\partial t}P^\prime_n=-\frac{v_0}{2}\frac{\partial}{\partial r}\left(\exp\left(-(|n+1|^\alpha-|n|^\alpha)\, \frac{t}{\tau}\right)P^\prime_{n+1}
+\exp\left(-(|n-1|^\alpha-|n|^\alpha)\, \frac{t}{\tau}\right)P^\prime_{n-1}\right)+\nonumber \\
&&-
\frac{n}{2}\left(\frac{v_0}{r}-\kappa \right)\left(\exp\left(-(|n+1|^\alpha-|n|^\alpha)\, \frac{t}{\tau}\right)P^\prime_{n+1}
-\exp\left(-(|n-1|^\alpha-|n|^\alpha)\, \frac{t}{\tau}\right)P^\prime_{n-1}\right)\,.
\label{eq:fpe_pnt}
\end{eqnarray}
Considering the index $n=0$, we note that for $n=0$ the components $P_0'$ and $P_0$ converge and equate to the marginal distance pdf $P(r,t)$. 

We take another partial time derivative in of $P_0^\prime$ in \eqref{eq:fpe_pnt}. In the obtained relation we replace  $P_1^\prime(t)$ using \eqref{eq:fpe_pnt} with $n=0$ at the l.h.s. It leads to an expression containing first and second derivatives of $P_0^\prime$ and terms with $P_{\pm 2}^\prime$:
\begin{eqnarray}
&&\frac{\partial}{\partial t^2}P^\prime_0+\frac{1}{\tau}\frac{\partial}{\partial t}P^\prime_0=
\frac{v_0^2}{2}\frac{\partial}{\partial r}\left(\frac{\partial}{\partial r}P^\prime_{0}-\left(\frac{1}{r}-\frac{\kappa}{v_0} \right)P^\prime_{0}\right)+\nonumber \\
&&+\frac{v_0^2}{4}\exp\left(-|2|^\alpha\, \frac{t}{\tau}\right)\frac{\partial}{\partial r}\left(\frac{\partial}{\partial r}\left(P^\prime_{2}+P^\prime_{-2}\right)
+\left(\frac{1}{r}-\frac{\kappa}{v_0} \right)\left(P^\prime_{2}-P^\prime_{-2}\right)\right)\,.
\label{eq:fpe_p0telep2}
\end{eqnarray}
For $t\gg\tau$ the terms containing $P_{\pm 2}$ vanish and as was discussed in \cite{sevilla_telegraph,Sevilla_smol} the second time derivative
containing a ballistic part of the motion can be also neglected if the limit of $v_0^2\tau$ for small $\tau$ remains finite. Therefore, we are left 
under these conditions with the overdamped description:
\begin{eqnarray}
&&\frac{\partial}{\partial t}P(r,t)=D_{\rm{eff}}\frac{\partial}{\partial r}\left(\frac{\partial}{\partial r}P(r,t)-\left(\frac{1}{r}-\frac{\kappa}{v_0} \right)P(r,t)\right)\,,
\label{eq:fpe_over}
\end{eqnarray}
with the effective diffusion coefficient 
\begin{equation}
D_{\rm{eff}}=\frac{v_0^2\tau}{2}\,.
\label{eq:deff_search1}
\end{equation}
The resulting Smoluchowski equation  determines the overdamped dynamics of the spatial transition pdf $P(r,t|r_0,t_0)$ as used in Section \ref{sec:tau}. Since its validity is bounded to time scales $t \gg \tau$,  the initial angle $z_0$ is forgotten and the  pdf does to depend on $z_0$, further on. Small $\tau$ implies large noise or small velocities in agreement with \eqref{eq:tau}.

\section{Mechanics of the searcher}
\label{app:celes}
Here we elaborate some similarities of the deterministic dynamics to the celestial mechanics. The motion of the particle reminds the planetary motion around a central body in a attracting potential. But at first we underline the main differences. In our problem we always deal with  constant speed of the particle which is another integral of motion in the problem.  

In consequence, the kinetic energy of the particle becomes (comp. Equation \eqref{eq:ekin})
\begin{equation}
  E_{kin}= \frac{1}{2}\left( \frac{{\rm d}r}{{\rm d}t}\right)^2 +v_0^2 \sin^2(z)= \frac{1}{2}v_0^2\,.
\end{equation}
After replacement of the $\sin(z)$-item by the constant integral of motion $X(r,z)$ the $z(t)$ variable disappears and the energy reads 
\begin{equation}
  E_{kin}= \frac{1}{2}\left( \frac{{\rm d}r}{{\rm d}t}\right)^2 + \frac{1}{2} v_0^2 \left(\frac{X}{r}\right)^2 \exp\left(2 \frac{r}{r_c}\right)= \frac{1}{2}v_0^2\,.
 \label{eq:eff_en}
\end{equation} 
This equation could be reinterpreted as describing the full mechanical
energy for a particle moving in a central force field at distance $r(t)$ and with effective
potential
\begin{equation}
  U_{\rm eff}(r)=\frac{1}{2} v_0^2\left(\frac{X}{r}\right)^2 \exp\left(2 \frac{r}{r_c}\right)\,.
  \label{eq:ueff}
\end{equation}  
The effective central force acting on the particle follows as
particle is
\begin{equation}
  F_{\rm eff}(r)= -\frac{\partial}{\partial r} U_{\rm eff}(r) = \left(\frac{X}{r}\right)^2 \left(\frac{1}{r}-\frac{1}{r_c}\right) \exp\left(2 \frac{r}{r_c}\right).
\end{equation} 
So, the motion of the particle is conservative and it moves through an effective central field, with the  potential energy given by \eqref{eq:ueff}. This effective potential becomes infinity at $r=0$ and if $r \rightarrow \infty$. Hence, the particle performs only bounded oscillatory motion with full energy $v_0^2/2$. At perihelion and aphelion with $r_{\rm max/min}$, respectively, the potential 
energy is extremal $U_{\rm{eff}}=1/2$, as the radial velocity vanishes at those distances. The force changes between attraction and repulsion at $r=r_c$ where the radial velocity of the particle is maximal.

\section*{References}

%

\end{document}